\documentclass[aps,prd,superscriptaddress,twocolumn]{revtex4-2}
\usepackage{graphicx}
\usepackage[caption=false]{subfig}
\usepackage{epstopdf}
\usepackage{amsmath}
\usepackage{amsfonts}
\usepackage{amssymb}
\usepackage{latexsym}
\usepackage{hyperref}
\usepackage[english]{babel}
\usepackage[utf8]{inputenc}
\usepackage[colorinlistoftodos]{todonotes}
\usepackage{color}
\usepackage{slashed}
\usepackage{feynmp}
\usepackage{bm}
\usepackage{bbold}
\usepackage{eufrak}
\usepackage{slashed}
\usepackage{bm}  % To make greek letters in bold face {\bm \Phi}
\usepackage{tabu}
\usepackage{xcolor}
\usepackage{tikz-feynman}

\begin{document}

\title{Lorentz-symmetry violation in the electroweak sector: scattering processes in future $e^+ \, e^-$ colliders}

\author{P. De Fabritiis}\email{pdf321@cbpf.br}
\affiliation{Centro Brasileiro de Pesquisas F\'{i}sicas (CBPF), Rua Dr Xavier Sigaud 150, Urca, Rio de Janeiro, Brazil, 22290-180}
	
	%\author{Y. M. P. Gomes}\email{ymuller@cbpf.br}
	%\affiliation{Universidade do Estado do Rio de Janeiro (UERJ), Rua São Francisco Xavier, 524, Maracanã, Rio de Janeiro, Brazil, CEP 20550-013}
	
\author{P. C. Malta}\email{pedrocmalta@gmail.com}
\affiliation{R. Antonio Vieira 23, 22010-100, Rio de Janeiro, Brazil}

\author{M. J. Neves}\email{mariojr@ufrrj.br}
\affiliation{Departamento de F\'isica, Universidade Federal Rural do Rio de Janeiro, BR 465-07, 23890-971, Serop\'edica, RJ, Brazil}

%%%%%%%%%%%%%%%%%%%

\begin{abstract}
We study CPT-odd non-minimal Lorentz-symmetry violating couplings in the electroweak sector modifying the interactions between leptons, gauge mediators and the Higgs boson. The tree-level (differential) cross sections for three important electroweak processes are discussed: $e^+ \, e^- \rightarrow Z \, H$, $e^+ \, e^- \rightarrow Z \, Z$ and $\gamma \, \gamma \rightarrow W^+ \, W^-$. By considering next-generation $e^+ \, e^-$ colliders reaching center-of-mass energies at the TeV scale and the estimated improved precision for the measurements of the respective cross sections, we are able to project upper bounds on the purely time-like background 4-vector as strict as $\lesssim 10^{-5}  \, \mbox{GeV}^{-1}$, in agreement with previous work on similar Lorentz-violating couplings.
\end{abstract}

\pacs{11.30.Cp, 12.60.-i, 13.35.Bv, 14.70.Fm, 14.70.Hp}
\maketitle

%%%%%%%%%%%%%%%%%%%%%%

\section{Introduction} \label{sec_intro}
	
\indent

The Standard Model (SM) of particle physics is certainly one of the most successful theories ever devised, explaining phenomena ranging from electromagnetic to nuclear interactions. Since its earliest days, more than fifty years ago, the
SM raked up victories culminating in the discovery of the Higgs boson in 2012, filling in an important missing piece of the particle physics puzzle. Despite of its many noteworthy achievements, there are several observations that cannot be described within its framework, {\it e.g.}, neutrino masses and dark matter/energy. This means that the SM cannot be a final theory and it is necessary to go beyond it to achieve a deeper understanding of the laws of Nature. Lorentz invariance is one of the cornerstones of the SM and has survived close scrutiny for almost a century. Nevertheless, there are still issues unexplained within the framework of the SM and even this well-founded symmetry must be systematically tested in high-precision measurements.

Historically, hadron and lepton colliders have played complementary roles in high-energy physics. Most discoveries were made in the former, with subsequent precision measurements performed in the latter. An example is the discovery of the W-bosons at CERN's SPS $p\overline{p}$ Collider \cite{ua1, ua2}. This discovery, in particular the determination of the masses of the W-boson, was later refined at LEP~\cite{aleph90}. Similarly, the next step in electroweak (EW) physics is to achieve even higher precision in the EW parameters associated with the Higgs and gauge bosons, a task ideally suited for $e^+ \, e^-$ colliders. As a matter of fact, in the last couple decades the high-energy community has debated the prospects for next-generation lepton machines and a few candidates turned out to have realistic chances of being built: Circular Electron Collider (CEPC, in China)~\cite{cepc}, Future Circular Collider (FCC, at CERN)~\cite{fcc1,fcc2}, Compact Linear Collider (CLIC, at CERN)~\cite{clic1} and International Linear Collider (ILC, prospectively in Japan)~\cite{ilc, white, ilc_TR}.

Besides the trade-offs between circular (CEPC, CLIC) and linear (FCC, ILC) accelerator designs~\cite{blondel}, it is clear that $e^+ \, e^-$ colliders would provide ideal grounds for precision tests of the EW theory, in particular of the Higgs and gauge sectors~\cite{tadeusz}. While confirming the SM predictions has been a relatively steady trend in collider experiments, increasingly precise measurements may point to possible discrepancies -- or rather, new physics -- which are long expected to be found~\cite{ellis}. A leading candidate to supplement the SM in ultra-high energy scales is (super)string theory, which could have potentially observable effects already at $\sim$~TeV scales~\cite{quevedo, gaillard}, such as the appearance of supersymmetric particles~\cite{drees, boubaa, tata} and axion-like particles~\cite{cicoli, coriano, axion}.

Interestingly enough, not only new particles are predicted by such extension scenarios. In string and quantum gravity models, for example, it is possible that Lorentz symmetry itself be broken at a very high energy scale once tensor fields acquire non-trivial vacuum expectation values and freeze, thus becoming time-independent. These tensors may single out directions in space-time, consequentially breaking Lorentz symmetry~\cite{Kost1, Posp1, Mavro1, Mavro2, Moffat, Camelia, ColemanPRD99}. Particularly relevant is the fact that these non-dynamical tensors -- backgrounds, actually -- are coupled to the SM fields, so that Lorentz-symmetry violating (LSV) effects would not be limited to unattainably high energy scales, but may play a role in low-energy phenomena. These couplings are systematically explored in the so-called Standard Model Extension (SME)~\cite{Colladay,Colladay2} and upper bounds on several of the LSV coefficients are collected in ref.~\cite{tables}; see also ref.~\cite{Mattingly}.

In this paper we introduce non-minimal LSV couplings in the EW sector of the SM. Modifications of this sector have been proposed in refs.~\cite{DColladay2018, DColladay2017Symmetry, DColladay2017, Lehnert}, where the modified dispersion relations of the photon, Z- and W- bosons were explored. Furthermore, non-minimal LSV couplings were considered in different scenarios within QED, {\it e.g.}, scattering processes, the spectrum of the hydrogen atom and electric or magnetic dipole moments~\cite{Belich1, Posp3, Posp4, Belich2, Gomes, scat_LSV, Bakke, Anacleto}. Also decays related to flavor-changing neutral currents and lepton flavor violation have been analyzed~\cite{YuriHelayel2020}. It should be emphasized that the SME is a more general setup which, by construction, encompasses the non-minimal extension that we propose in the following.

Here we generalize the prescription from ref.~\cite{scat_LSV} by introducing two LSV 4-vectors, $\rho$ and $\xi$, related to the $SU(2)_{\rm L}$ and $U(1)_{\rm Y}$ sectors, respectively; this scenario is similar to that analyzed in refs.~\cite{mouchrek1, mouchrek2, Mario}. As a consequence, after spontaneous symmetry breaking, new couplings of the W- and Z-bosons among themselves and the Higgs boson, as well as with the leptons, arise. In ref.~\cite{Mario} we employed a similar prescription, but kept $\rho \neq \xi$, focusing on the interaction vertices involving gauge bosons and leptons, and obtaining bounds on the LSV parameters through the analysis of tree-level decay rates and branching ratios. Here, in contrast, we single out the case $\rho = \xi$ and study scattering processes relevant to future $e^{+}\,e^{-}$ colliders, where EW parameters will be measured with unprecedented precision. By obtaining the tree-level cross sections for the processes $e^+ \, e^- \rightarrow Z \, H$, $e^+ \, e^- \rightarrow Z \, Z $ and $\gamma \, \gamma \rightarrow W^+ \, W^-$ including the LSV-induced corrections and comparing these with projected experimental sensitivities, we are able to place upper bounds (projections) on the LSV parameters, in particular on the time-like component.

%%%%% SCF %%%%%%%

As a final and mandatory remark, we would like to stress a point regarding the components of the LSV 4-vector as measured at an Earth-based laboratory. The background $\rho$ is, by construction, non-dynamical, being fixed in space-time. However, this is manifest only in a truly inertial reference frame. A common and convenient choice is the so-called Sun-centered frame (SCF)~\cite{ref_sun}, which may be considered approximately inertial for our purposes. Consequently, in Earth-bound experiments, the LSV-modified observables, such as cross sections, should display sidereal time modulations that could be picked up by carefully time-stamping the data.

It is thus necessary to express the laboratory-frame components, $\rho^{\mu}_{\rm lab}$, in terms of the components, $\rho_{\rm SCF}^{\mu}$, which are static in the SCF. Following ref.~\cite{ref_sun}, we find that, up to first order in Earth's orbital velocity ($\sim 10^{-4}$ in units of $c$), $\rho_{\rm lab}^{0} \simeq \rho_{\rm SCF}^{T}$ and ${\bm \rho}_{\rm lab}^{i} \simeq R^{iJ} (\chi,T) {\bm \rho}_{\rm SCF}^{J}$, where $R^{iJ} (\chi,T)$ is a time-dependent rotation matrix; see also ref.~\cite{Mario}. In the context of the approximations above, the time component is easily factored out from the squared amplitudes. The situation with the spatial components is substantially more complex, so that, for the sake of simplicity, we shall henceforth assume that the LSV background is purely time-like at the SCF.

%%% org

This paper is organized as follows: in sec.~\ref{sec_model} we present our implementation of LSV in the EW sector of the SM. In sec.~\ref{sec_higgs} we discuss the Higgs sector and the consequences of our LSV prescription to the Higgs, gauge and lepton sectors after spontaneous symmetry breaking. Next, in sec.~\ref{sec_app} we explore a few interesting scattering processes that are affected by LSV. Finally, in sec.~\ref{sec_conclusion} we present our closing remarks. We use natural units ($c = \hbar = 1$) and the flat Minkowski metric $\eta^{\mu\nu}={\rm diag}(+1,-1,-1,-1)$ throughout.

%Finally, in section~\ref{sec_conclusion} we present our concluding remarks. In our calculations we have used the Mathematica Package-X~\cite{PackageX} to automatically evaluate the contractions from the polarization and spin averaging procedures.

%%%%%%%%%%%%%%%%%%%%%%%%
	
\section{Theoretical structure of our LSV prescription} \label{sec_model}
\indent

Let us consider the EW Lagrangian
\begin{equation}
\mathcal{L}_{\rm EW}  =  \mathcal{L}_{\rm Higgs} + \mathcal{L}_{\rm Gauge} + \mathcal{L}_{\rm Leptons} \; ,
\end{equation}
with
\begin{subequations}
\begin{eqnarray} \label{lags}
\mathcal{L}_{\rm Higgs} & = & \vert D_\mu H \vert^2 - V\left(H\right) \; ,
\\
\mathcal{L}_{\rm Gauge} & = & - \frac{1}{4} \, F_{\mu \nu}^{\;\;\;\,a} \, F^{\mu\nu a} - \frac{1}{4} \, B_{\mu \nu }\, B^{\mu \nu } \; ,
\\
\mathcal{L}_{\rm Leptons} & = & i \, \bar{L}_{i} \gamma^\mu D_\mu L_{i} + i \, \bar{\ell}_{iR} \gamma^\mu D_\mu \ell_{iR} \label{lag_fer}\; ,
\end{eqnarray}
\end{subequations}
where the fermion content is
\begin{equation}
L_{i} = \left( \nu_{\rm iL} \,\,\, \ell_{\rm iL} \right)^t \quad {\rm and} \quad \ell_{\rm iR},
\end{equation}
with $i=\left\{e, \mu, \tau \right\}$; the hypercharge assignments are: $Y(L_{i}) = -1/2$ and $Y(\ell_{iR}) = -1$. The fields $A_\mu^{\;\,a}$, with $a=\left\{1,2,3 \right\}$, and $B_\mu$ are the $SU(2)_L$ and $U(1)_Y$ gauge fields, respectively. Their dynamics is described with the help of the field-strength tensors $F_{\mu\nu}^{\;\;\;\,a} = \partial_\mu A_\nu^{\;\,a} - \partial_\nu A_\mu^{\;\,a} + g \, f^{a b c} \, A_\mu^{\;\,b} \, A_\nu^{\;\,c}$ and $B_{\mu \nu } = \partial_\mu B_\nu - \partial_\nu B_\mu$. In the scalar sector, $H$ is the Higgs doublet, a complex scalar field whose potential is given by
\begin{equation}\label{h_pot}
V\left(H\right) = \lambda \left( H^\dagger H - \frac{\mu^2}{  2 \lambda} \right)^2 \; ,	
\end{equation}
where $\mu^2$ and $\lambda$ are real and positive parameters determining the shape of the potential.

The covariant derivative with respect to the EW gauge group $SU(2)_L \times U(1)_Y$ is defined by
\begin{equation}\label{cov_dev}
D_\mu = \partial_\mu  - i \, g \, A_\mu^{\;\,a} \, T^a - i \, g' \, Y \, B_\mu \; .
\end{equation}
Here $T^a$ are the Hermitean generators of $SU(2)$ satisfying $\left[  T^a  , T^b \right] = i \, f^{a b c} \, T^c$, and $Y$ is the associated hypercharge. The gauge couplings $g$ and $g'$ are associated with $SU(2)_L$ and $U(1)_Y$, respectively. The commutator of the covariant derivative gives
\begin{equation}
\left[ \, D_\mu \, , \, D_\nu \, \right] = -i \, g \, F_{\mu \nu}^{\;\;\;\,a} \, T^a - i \, g' \, Y \, B_{\mu \nu} \; ,
\end{equation}
and we have the Bianchi identities
\begin{subequations}
\begin{eqnarray}\label{bianchi_1}
D_\mu \tilde{F}^{\mu \nu} &=& 0 \; ,
\\
D_\mu \tilde{B}^{\mu \nu} &=& 0 \; ,
\end{eqnarray}
\end{subequations}
where the dual field-strength tensor is defined as $\tilde{F}^{\mu \nu} = \frac{1}{2} \epsilon^{\mu \nu \rho \sigma} F_{\rho \sigma}^{\;\;\;\,a} \, T^a$, and similarly for $\tilde{B}_{\mu \nu}$. Here $\epsilon^{\mu \nu \rho \sigma}$ is the totally antisymmetric tensor with $\epsilon^{0 1 2 3} = + 1$.

Let us consider the following LSV prescription
\begin{equation} \label{cov_dev_lsv}
D_\mu \rightarrow \hat{D}_\mu = D_{\mu} + i \rho^\nu \, g \, F_{\mu \nu}^{\;\;\;\,a} \, T^a
+ i \, \xi^\nu g^{\prime} \, Y B_{\mu \nu} \; ,
\end{equation}
where $\rho$ and $\xi$ are real and constant LSV 4-vectors with canonical dimension of inverse mass. This prescription introduces non-minimal Lorentz-violating operators in the Higgs sector in a way consistent with the usual gauge transformations, thus preserving the whole gauge structure of the SM. Since LSV has not been observed, we may assume that these backgrounds are small relative to currently attainable energy scales, so that we are allowed to keep only terms at leading order in the LSV parameters, and consistently neglect all higher-order terms. Our LSV prescription introduces non-renormalizable operators and should be understood as an effective theory. We finally remark that extensions of the covariant derivative including non-minimal couplings with a LSV background were already studied before; the interested reader should refer to refs.~\cite{YuriHelayel2020, mouchrek1, mouchrek, Belich3} and references therein.

Given the LSV prescription, one can immediately show that, if we take into consideration the usual gauge transformations for the gauge fields, this extended covariant derivative will have the same transformation properties under gauge transformations as the usual one, that is,
\begin{equation}\label{key}
 H' = U H  \implies (\hat{D}_\mu H)' = U \, (\hat{D}_\mu H) \; .
\end{equation}
On the other hand, we can also see that, imposing the above equation and understanding the prescription as an expansion in the LSV parameters, by comparing both sides of the equation we can obtain the usual gauge transformations for the gauge fields. Therefore, the extended derivative is covariant in the usual sense. The commutator of these objects reads
\begin{equation}\label{ComutD}
\left[ \, \hat{D}_\mu \, , \, \hat{D}_\nu \, \right] = - i g \, \hat{F}_{\mu \nu}-ig' \, Y \, \hat{B}_{\mu\nu} \; ,
\end{equation}	
where we defined
\begin{subequations}
\begin{eqnarray}\label{key}
\hat{F}_{\mu \nu} &=& F_{\mu \nu} +  \rho^\beta \, \left[ \, D_\beta \, , \,  F_{\mu \nu} \, \right]
\nonumber \\
&=& \left(F_{\mu \nu}^{\;\;\;\,a} + \rho^\beta \partial_\beta F_{\mu \nu}^{\;\;\;\,a} + \rho^\beta g \, f^{a b c}
A_\beta^{\;\,\,b} \, F_{\mu \nu}^{\;\;\;\,c}\right) T^a \, ,
\hspace{1cm}
\\
\hat{B}_{\mu \nu} &=& B_{\mu \nu} +  \xi^\beta \left[   D_\beta,  B_{\mu \nu} \right]  \nonumber \\
&=& B_{\mu \nu} + \xi^\beta \partial_\beta B_{\mu \nu} \; .
\end{eqnarray}
\end{subequations}
These objects should be understood as an extended version of the field-strength tensors. In this case, we have to investigate what is the effect of acting with the extended covariant derivative in these extended field-strength tensors to see whether the usual gauge structure is maintained or not. By doing that, we obtain
\begin{subequations}
\begin{eqnarray}\label{IdBianchiG}
\hat{D}_\mu \tilde{\hat{F}}^{\mu\nu} &=& -ig' \, Y \left( \rho^{\beta}-\xi^{\beta} \right) B_{\mu\beta}\tilde{F}^{\mu\nu} \; ,
\\
\hat{D}_\mu \tilde{\hat{B}}^{\mu\nu} &=& -ig \left( \xi^{\beta}-\rho^{\beta} \right) F_{\mu\beta}\tilde{B}^{\mu\nu} \; .
\end{eqnarray}
\end{subequations}

Considering the more natural scenario where the background LSV vectors couple in an equivalent way with the different sectors of the gauge group, that is, $\rho= \xi$, we immediately see that the above equations result in extended versions of the Bianchi identities. In most of what follows, we will adopt this setup, but we will keep track of the terms proportional to $\left( \xi-\rho \right)$ to investigate what could be the consequences of having a very small, but non-vanishing, value for the difference between the two LSV backgrounds.

%$\hat{D}_\mu \tilde{\hat{F}}^{\mu\nu} = \hat{D}_\mu \tilde{\hat{B}}^{\mu\nu} = 0$

Following our prescription for the covariant derivative, it is also natural to extend the field-strength tensors stemming from their commutator. Therefore, let us consider the following extension for the gauge kinetic term
\begin{equation}\label{key}
\mathcal{L}_{\rm Gauge} \rightarrow \hat{\mathcal{L}}_{\rm Gauge} = - \frac{1}{4} \, \hat{F}_{\mu \nu}^{\;\;\;\,a} \, \hat{F}^{\mu \nu a}
- \frac{1}{4} \, \hat{B}_{\mu \nu} \hat{B}^{\mu \nu} \; .
\end{equation}
Using the expressions obtained for the extended field-strength tensors, one sees that, since $\rho^\nu$ and $\xi^\nu$ are real constant LSV parameters, this extended kinetic terms will differ from the usual ones only by a total derivative that can be considered as a boundary term in the action, and therefore can be consistently neglected. Thus, even if we are led to consider extended field-strength tensors to construct our gauge kinetic terms, this modification does not bring any new term in the pure gauge sector; new contributions arise only in the Higgs and fermion sectors, as we will see shortly. The same prescription must be applied to the fermionic sector, so that eq.~\eqref{lag_fer} is modified to read
\begin{equation}\label{Lleptonschapeu}
\mathcal{L}_{\rm Leptons} \rightarrow \hat{\mathcal{L}}_{\rm Leptons} = i \bar{L}_{i} \gamma^\mu \hat{D}_\mu L_{i} + i \bar{\ell}_{iR} \gamma^\mu \hat{D}_\mu \ell_{iR} \; .
\end{equation}
In fact, as will be discussed in the next section, this LSV prescription will not change the quadratic sector as a whole, but as we identify the physical degrees of freedom and rewrite the Lagrangian in terms of these fields, we will observe the appearance of LSV terms in the pure gauge sector, as well as the already expected modifications in the Higgs and fermion sectors. Therefore, we conclude that the LSV backgrounds will manifest themselves only through interactions.

%%%%%%%%%%%%%%%%%%%%%%%%%%%%%%%%5

\section{The LSV-modified EW sector} \label{sec_higgs}

We first consider the Higgs sector of the EW Lagrangian. As is well-known, the Higgs potential induces a non-trivial vacuum expectation value for the Higgs field, given by $\vert \langle H \rangle \vert^2 = \frac{v^2}{2} = \frac{\mu^2}{2 \lambda}$, where the Fermi scale is fixed experimentally to be $v \approx 246 \, \text{GeV}$ through the beta decay of the muon. At this stage we have electroweak symmetry breaking (EWSB), putting the theory into the Higgs phase with three massive vector bosons $W^+, W^-, Z$, a massive scalar $h$ and the massless photon $A$ in the spectrum.

Expanding around $v$ and choosing the unitary gauge, we can write the Higgs isodoublet as
\begin{equation}
H(x) = \frac{1}{\sqrt{2}} \left( \begin{array}{c}
0
\\
\\		
v + h(x)
\end{array} \right) \; ,
\end{equation}
whose hypercharge assignment is $Y(H) = +1/2$. Given that our prescription does not alter the quadratic part of the gauge Lagrangian, we can define the physical intermediate bosons as usual:
\begin{subequations}
\begin{eqnarray}
W^{\pm}_\mu &=& \frac{1}{\sqrt{2}}\left( A_\mu^1 \mp i A_\mu^2 \right) \; ,
\\
Z_\mu &=& \cos\theta_W \, A_\mu^3 - \sin\theta_W \, B_\mu \; ,
\\
A_\mu &=& \sin\theta_W \, A_\mu^3 + \cos\theta_W \, B_\mu \; ,
\end{eqnarray}
\end{subequations}
where $\theta_W$ is the Weinberg angle, experimentally determined as $\sin^2\theta_W = 0.23$. The SM Higgs sector is then
\begin{eqnarray}\label{higgs_usual}
\mathcal{L}_{\rm Higgs}  &=& \frac{1}{2} \left(\partial_\mu h\right)^2 - \frac{1}{2} \, m_h^2 \, h^2
- \frac{gm_h^2}{4 m_W} \, h^3 - \frac{g^2 m_h^2}{32 m_W^2} \, h^4 \nonumber \\
%\hspace{-0.5cm}
&+& \left( 1 + \frac{h}{v} \right)^2 \left[m_W^2 W_\mu^+ W^{\mu-}  + \frac{1}{2} \, m_Z^2 Z_\mu Z^\mu \right] \! ,
\hspace{0.5cm}
\end{eqnarray}
where  $m_W = gv/2 = 80.4 \, \text{GeV}$, $m_Z = m_W/\cos\theta_W = 91.2 \, \text{GeV}$,
$m_h = v \sqrt{2 \lambda} = 125.1 \, \text{GeV}$. As we will see, these results will remain unaffected
by the LSV prescription discussed here.

Let us now consider the Higgs sector with the LSV-extended covariant derivative, that is
\begin{equation}\label{key}
\mathcal{L}_{\rm Higgs} \rightarrow \hat{\mathcal{L}}_{\rm Higgs} = \vert \hat{D}_\mu H \vert^2 - V(H) \; .
\end{equation}
Here we have, besides the terms in eq.~\eqref{higgs_usual}, the following LSV contributions
\begin{eqnarray}\label{HiggsLorentz}
\vert \hat{D}_\mu H \vert^2 & \supset &  \left( 1 + \frac{h}{v} \right)^2 \! \bigg[ - m_W^2 \rho^\nu \left(W^+_\mu W^-_{\mu \nu} + W^-_\mu W^+_{\mu \nu}\right)  \nonumber \\
&-& i e m_W^2 \rho^\nu A^\mu \left(W_\mu^+ W_\nu^- - W_\mu^- W_\nu^+\right)
\nonumber\\
&+& i e m_W^2 \tan\theta_W \rho^\nu \, Z^\mu \left(W_\mu^+ W_\nu^- - W_\mu^- W_\nu^+\right)
\nonumber \\
&-& m_Z^2 \left(\sin^2\theta_W \xi_\nu + \cos^2\theta_W \rho_\nu \right) Z_\mu Z^{\mu \nu}
\nonumber \\
&+& m_Z^2 \sin\theta_W \cos\theta_W \left(\xi_\nu - \rho_\nu\right) Z_\mu \, F^{\mu \nu}  \bigg] \; ,
\end{eqnarray}
where we omitted the SM term $\vert D_\mu H \vert^2$ from the equation above for the sake of clarity.

The gauge kinetic part will be the usual one, since the extension of the field-strength tensors does not introduce new terms up to total derivatives, as we already discussed. After EWSB, omitting the interaction terms, the gauge kinetic Lagrangian may be written as
\begin{equation}\label{gaugekin}
\mathcal{L}_{\rm Gauge}^{\rm Kin} = - \frac{1}{4} \, F_{\mu \nu} F^{\mu \nu} - \frac{1}{4} \, Z_{\mu \nu} Z^{\mu \nu}
- \frac{1}{2} \, W_{\mu \nu }^+ W^{\mu \nu -} \, ,
\end{equation}
where, as usual, we have $F_{\mu \nu} = \partial_\mu A_\nu - \partial_\nu A_\mu$, $Z_{\mu \nu} = \partial_\mu Z_\nu - \partial_\nu Z_\mu$, and $W_{\mu \nu}^{\pm} = \partial_\mu W_\nu^\pm - \partial_\nu W_\mu^\pm$. Therefore, we are investigating here the total Lagrangian
\begin{equation}\label{L_ew}
\hat{\mathcal{L}}_{\rm Bosons} = \hat{\mathcal{L}}_{\rm Higgs} + \hat{\mathcal{L}}_{\rm Gauge} \; .
\end{equation}
Given the Lagrangian above, we can obtain the equations of motion for the gauge fields, which, upon omission of the interactions, read
\begin{subequations}
\begin{eqnarray}
\partial_{\mu}W^{\mu\nu} \! &+& \! m_{W}^2W^{\nu}\!-\!m_{W}^2 \rho_{\mu}\partial^{\nu}W^{\mu}\!+\!m_{W}^2\rho^{\nu}\,\partial_{\mu}W^{\mu} = 0 \, ,
\hspace{0.78cm}
\\
\partial_{\mu}Z^{\mu\nu} \!&+&\! m_{Z}^2\,Z^{\nu}\!-\!m_{Z}^2\sin\theta_W \cos\theta_W\left(\xi_{\mu}-\rho_{\mu}\right)F^{\mu\nu}
\nonumber \\
&+& m_{Z}^2\left(\sin^2\theta_{W}\xi^{\nu}+\cos^2\theta_{W}\rho^{\nu} \right)\partial_{\mu}Z^{\mu}
\nonumber \\
&-& m_{Z}^2 \left(\sin^2\theta_{W}\xi_{\mu}+\cos^2\theta_{W}\rho_{\mu} \right) \partial^{\nu}Z^{\mu} = 0 \; ,
\\
\partial_{\mu}F^{\mu\nu} \!&-&\! m_{Z}^2\sin\theta_W \cos\theta_W\left(\xi^{\nu}-\rho^{\nu}\right)\partial_{\mu}Z^{\mu}
\nonumber \\
& + & m_{Z}^2\sin\theta_W \cos\theta_W\left(\xi^{\mu}-\rho^{\mu}\right)\partial_{\mu}Z^{\nu}=0 \; .
\end{eqnarray}
\end{subequations}
If we apply a partial derivative to the equations above, we obtain the subsidiary conditions
\begin{subequations}
\begin{eqnarray}\label{sub_cond}
\partial_{\mu}W^{\mu} &=& \Box\left(\rho \cdot W\right) \; ,
\\
\partial_{\mu}Z^{\mu} &=& \left( \sin^2\theta_W \xi_\mu + \cos^2\theta_W \rho_\mu \right) \Box Z^\mu \; .
\end{eqnarray}
\end{subequations}
Plugging them back into the equations of motion and choosing the Lorenz gauge for the photon field, $\partial_\mu A^\mu =0$, we find
\begin{subequations}
\begin{align}
&\left(\Box+m_{W}^2\right) \left[W^\nu - \rho_\mu \partial^\nu W^\mu\right] = 0 \; ,
\\
&\left(\Box+m_{Z}^2\right) \left[\eta^{\mu\nu} - \partial^\nu \left(\sin^2\theta_{W}\xi_{\mu}+\cos^2\theta_{W}\rho_{\mu} \right) \right] Z^{\mu}
\nonumber \\
&- m_{Z}^2\sin\theta_{W}\cos\theta_{W} \left(\xi_{\mu}-\rho_{\mu}\right)F^{\mu\nu} = 0 \; ,
\\
&\Box A^{\nu} +m_{Z}^2\sin\theta_W \cos\theta_W\left(\xi^{\mu}-\rho^{\mu}\right)\partial_{\mu}Z^{\nu} = 0 \; .
\end{align}
\end{subequations}
It is worthwhile noting how the difference of the LSV backgrounds is responsible for mixing the Z-boson and the photon. However, we are primarily interested in a more natural scenario, where $\rho = \xi$, which allows the Bianchi identities to be satisfied. Therefore, rewriting the expressions above for this case of interest, we have
\begin{subequations}
\begin{eqnarray}
\partial_{\mu}W^{\mu\nu} &+& m_{W}^2(\eta^{\mu \nu} +  \rho^{\nu}\,\partial_{\mu}- \rho_{\mu}\partial^{\nu})W^{\mu} =0 \; ,
\\
\partial_{\mu}Z^{\mu\nu} &+& m_{Z}^2(\eta^{\mu \nu} + \rho^\nu \partial_{\mu} - \rho_\mu \partial^{\nu}   )Z^{\mu} = 0  \; ,
\\
\partial_{\mu}F^{\mu\nu} &=& 0 \; ,
\end{eqnarray}
\end{subequations}
where the decoupling of the Z-boson and the photon is made explicit. Incidentally, we note that the equations of motion of the photon are unaffected by the presence of the LSV background.

The subsidiary conditions are then reduced to
\begin{subequations}
\begin{eqnarray}\label{condsubW}
\partial_{\mu}W^{\mu} &=& \Box\left(\rho \cdot W\right) \; ,
\\
\partial_{\mu}Z^{\mu} &=& \Box \left( \rho \cdot Z \right) \; ,
\end{eqnarray}
\end{subequations}
which suggest a redefinition of the massive Z- and W-fields as
\begin{subequations}
\begin{eqnarray}
\check{W}^{\mu} &\equiv& W^{\mu}-\partial^{\mu}(\rho\cdot W) \; ,  \label{physW}
\\
\check{Z}^{\mu} &\equiv& Z^{\mu} - \partial^{\mu}(\rho\cdot Z) \; , \label{physZ}
\end{eqnarray}
\end{subequations}
so that the new fields will, by construction, satisfy the usual subsidiary conditions $\partial_{\mu}\check{W}^{\mu\pm}=0$, $\partial_{\mu}\check{Z}^{\mu}=0$. Furthermore, the equations of motion become
\begin{subequations}
\begin{eqnarray}
\left(\Box+m_{W}^2\right)\check{W}^{\mu} &=& 0 \; ,
\\
\left(\Box+m_{Z}^2\right)\check{Z}^{\mu} &=& 0 \; .
\end{eqnarray}
\end{subequations}
The redefined boson fields satisfy the standard Klein-Gordon equations. Keeping in mind that we have so far retained only quadratic terms, we see that the dispersion relations are not modified by the LSV background. Finally, the photon field does not need a redefinition and already describes the corresponding physical particle in this LSV prescription.

The next step is to re-write eq.~\eqref{L_ew} using the redefined fields via eqs.~\eqref{physW} and~\eqref{physZ}. All terms containing gauge bosons will be affected, independent of whether LSV was initially present or not. Let us first focus on the Higgs sector, as given in eq.~\eqref{higgs_usual}. The first line will remain untouched, but the second one will be modified. Limiting ourselves to leading order in the LSV parameters, we find that the quadratic terms will be reproduced, since the extra LSV terms can be written as innocuous boundary terms, which we then discard. Finally, also taking eq.~\eqref{HiggsLorentz} into account, the full Higgs sector Lagrangian, to leading order in LSV, can be written as
\begin{eqnarray} \label{L_higgs_lsv}
\hat{\mathcal{L}}_{\rm Higgs}   &=& \frac{1}{2} \left(\partial_\mu h\right)^2  - \frac{1}{2}\, m_h^2 \, h^2 - \frac{gm_h^2}{4 m_W} \, h^3 - \frac{g^2 m_h^2}{32 m_W^2} \, h^4
\nonumber \\
&&
\hspace{-1cm}
+\left( 1 + \frac{h}{v} \right)^2 \! \bigg[ \,  m_W^2 \, \check{W}^+_\mu \check{W}^{\mu-}  + \frac{1}{2} \, m_Z^2 \, \check{Z}_\mu \check{Z}^\mu
\nonumber \\
&&
\hspace{-1cm}
+ m_W^2 \rho^\nu \partial_\nu \! \left( \check{W}_{\mu}^{+} \check{W}^{\mu-} \right) + \frac{1}{2} \, m_Z^2 \, \rho^\nu \partial_\nu \left( \check{Z}_\mu \check{Z}^\mu \right)
\nonumber \\
&&
\hspace{-1cm}
-i e m_W^2 \rho_\nu \, A_\mu (\check{W}^{\mu+} \check{W}^{\nu-} - \check{W}^{\mu-} \check{W}^{\nu+} )
\nonumber \\
&&
\hspace{-1cm}
+i e m_W^2 \tan\theta_W \, \rho_\nu \, \check{Z}_\mu (\check{W}^{\mu+} \check{W}^{\nu-} - \check{W}^{\mu-} \check{W}^{\nu+} ) \!  \bigg].
\end{eqnarray}

We note that the third line above involves an apparent total derivative. This is true for the higgsless term in $\left( 1 + h/v \right)^2 $, but not for the other terms involving $h$ and $h^2$. Therefore, the quadratic part is left intact and the third line only modifies the usual W-Higgs and Z-Higgs vertices. We remark that the last two lines have a pure gauge contribution modifying the $\gamma WW$ and $ZWW$ vertices. Particularly noteworthy are the new dimension 4 and 5 operators that are induced by our LSV prescription considering the Higgs interactions with gauge bosons that can be read from the last two lines.
%5-field term $hh\gamma WW$, which is not present in the SM, thus being a distinctive feature of our LSV prescription.

Let us now move on to the pure gauge sector, which reads
\begin{equation}\label{key}
\hat{\mathcal{L}}_{\rm Gauge} = \mathcal{L}_{\rm Gauge}^{\rm Kin} + \mathcal{L}^{(3)}_{\rm Gauge} + \mathcal{L}^{(4)}_{\rm Gauge} \; ,
\end{equation}
where the kinetic part is shown in eq.~\eqref{gaugekin}. For the sake of simplicity, we quote here the triple and quartic parts of $\hat{\mathcal{L}}_{\rm Gauge}$ as in the SM before the field redefinitions:
\begin{eqnarray}\label{key}
&&\mathcal{L}^{(3)}_{\rm Gauge} = ie \! \left[ F_{\mu \nu} W^{\mu+} W^{\nu-}  \! - \!  A_\mu ( W^{\mu \nu+}  W_\nu^- \! - \! W^{\mu \nu-}  W_\nu^+ ) \right] \nonumber \\
&& + i e \cot\!\theta_{W} \! \left[ Z_{\mu \nu} W^{\mu+} W^{\nu-} - Z_{\mu}( W^{\mu \nu+}  W_\nu^- - W^{\mu \nu-}  W_\nu^+ ) \! \right],  \nonumber \\
\end{eqnarray}
\begin{eqnarray}
&&\mathcal{L}^{(4)}_{\rm Gauge} = e^2 \left( A_\mu W^{\mu+} A_\nu W^{\nu-} - A_\mu A^\mu \, W_\nu^+ W^{\nu-} \right) \nonumber \\
&& +\frac{e^2}{2\sin^2\theta_{W}} \left( W_\mu^+ W^{\mu+} W_\nu^- W^{\nu-} - W_\mu^+ W^{\mu-} W_\nu^+ W^{\nu-} \right)
\nonumber \\
&&+ e^2 \cot^2\theta_{W} \big( Z_\mu W^{\mu+} Z_\nu W^{\nu-} - Z_\mu Z^\mu W_\nu^+ W^{\nu-}
\nonumber \\
&& + W_\mu^+ A^{\mu} W_\nu^- Z^\nu \! + \! W_\mu^- A^{\mu} W_\nu^+ Z^{\nu} \! - \! 2 W_\mu^+ W^{\mu-} \! A_\nu Z^{\nu}  \big) \; .
\hspace{0.6cm}
\end{eqnarray}
The equations above will be modified once we redefine the gauge fields as per eqs.~\eqref{physW} and~\eqref{physZ}. We refrain from writing these expressions explicitly, as this would unnecessarily clutter the text; in sec.~\ref{sec_app} we will quote the relevant results as they are needed for each process. Please keep in mind that some pure gauge LSV terms are contained in eq.~\eqref{L_higgs_lsv}.

%%% leptons

The discussion so far was focused on the bosonic sector of the SM, but we need to address the effects of our LSV prescription  on the matter sector. Similarly to the gauge sector, our LSV prescription will not alter the kinetic sector, thus keeping the usual dispersion relations intact. The leptonic sector, cf. eq.~\eqref{Lleptonschapeu}, can be written as
\begin{align}\label{key}
\hat{\mathcal{L}}_{\rm Leptons} = \mathcal{L}_{\rm Leptons}^{\rm SM} + \hat{\mathcal{L}}_{\rm Leptons}^{(3)} + \hat{\mathcal{L}}_{\rm Leptons}^{(4)} \; ,
\end{align}
where
\begin{eqnarray}\label{LSMlep}
\mathcal{L}_{\rm Leptons}^{\rm SM}
%&= i \bar{L} \gamma^\mu D_\mu L + i \bar{R} \gamma^\mu D_\mu R  \nonumber \\
& = & i \, \bar{\ell}_{i} \gamma^\mu \partial_\mu \ell_{i}
+ i \, \bar{\nu}_{iL} \gamma^\mu \partial_\mu \nu_{iL} - e \, \bar{\ell}_{i} \gamma_{\mu} \ell_{i} \, A^{\mu}
\nonumber \\
&&
\hspace{-0.5cm}
+ \frac{g}{\sqrt{2}} \left( \bar{\nu}_{iL} \gamma^\mu \ell_{iL} \, \check{W}_\mu^+ + \bar{\ell}_{iL} \gamma^\mu \nu_{iL} \, \check{W}_\mu^{-}   \right)
\nonumber \\
&&
\hspace{-0.5cm}
+ \frac{g}{4 \cos\theta_W} \, \bar{\ell}_{i} \gamma_{\mu} \left( c_v + c_a \gamma^5  \right) \ell_{i} \, \check{Z}^{\mu}
\nonumber \\
&&
\hspace{-0.5cm}
+ \frac{g}{2 \cos\theta_W} \, \bar{\nu}_{iL} \gamma_{\mu} \nu_{iL} \, \check{Z}^{\mu}
\end{eqnarray}
describes the usual SM couplings of leptons and gauge bosons, and we defined
\begin{equation}  \label{c_va}
c_v = -1 + 4 \sin^2\theta_W \quad \quad \text{and} \quad \quad  c_a = 1 \, .
\end{equation}
The other Lagrangians comprise triple and quartic interactions containing the LSV background, already taking into account the field redefinitions. Explicitly, we have:
\begin{eqnarray}\label{L3_leptons_lsv}
\hat{\mathcal{L}}_{\rm Leptons}^{(3)} &=&
 \frac{g}{\sqrt{2}} \rho^{\nu}  \left[  \left( \bar{\nu}_{iL} \gamma^\mu \ell_{iL} \right) \partial_\nu \check{W}_{\mu}^+ +  \left( \bar{\ell}_{iL} \gamma^\mu \nu_{iL} \right) \partial_\nu \check{W}_{\mu}^- \right]
\nonumber \\
&&
\hspace{-1cm}
+ \frac{g}{2 \cos\theta_W} \, \rho^{\nu} \left( \bar{\nu}_{iL} \gamma^{\mu} \nu_{iL} \right)  \partial_\nu \check{Z}_{\mu}
%\nonumber \\
%&&
%\hspace{-0.8cm}
+ e \, \rho^{\nu} \left(\bar{\ell}_{i} \gamma^\mu \ell_{i}\right)  F_{\mu \nu}
%\right.
\nonumber \\
&&
\hspace{-1cm}
%\left.
+ \frac{g}{4 \cos\theta_W} \, \rho^{\nu}  \, \bar{\ell}_{i} \gamma^\mu \left( c_v + c_a \gamma^5 \right) \ell_{i} \, \partial_\nu \check{Z}_{\mu}	
\end{eqnarray}
and
\begin{eqnarray}\label{L4_leptons_lsv}
\hat{\mathcal{L}}_{\rm Leptons}^{(4)} &=& - \frac{i g^2}{2} \, \rho^\nu \left( \check{W}_\mu^+ \check{W}_\nu^- - \check{W}_\mu^- \check{W}_\nu^+ \right) \bar{\ell}_{iL} \gamma^\mu \ell_{iL}
\nonumber  \\
&&
\hspace{-1cm}
+ \frac{i g^2}{2} \, \rho^\nu \left( \check{W}_\mu^+ \check{W}_\nu^- - \check{W}_\mu^- \check{W}_\nu^+ \right) \bar{\nu}_{iL} \gamma^\mu \nu_{iL}
\nonumber  \\
&&
\hspace{-1cm}
- ig^2 \sin\theta_{W} \rho^\nu \left( \check{W}_\mu^+ A_\nu - \check{W}_\nu^+ A_\mu \right) \left(\bar{\nu}_{iL} \gamma^\mu \ell_{iL}\right)
\nonumber \\
&&
\hspace{-1cm}
+ i g^2 \, \sin\theta_{W} \rho^\nu \left( \check{W}_\mu^- A_\nu - \check{W}_\nu^-  A_\mu \right) \left(\bar{\ell}_{iL} \gamma^\mu \nu_{iL}\right)
\nonumber  \\
&&
\hspace{-1cm}
- ig^2 \, \cos\theta_{W} \rho^\nu  \left( \check{W}_\mu^+  \check{Z}_\nu - \check{W}_\nu^+ \check{Z}_\mu \right) \left(\bar{\nu}_{iL} \gamma^\mu \ell_{iL} \right)
\nonumber \\
&&
\hspace{-1cm}
+ i g^2 \, \cos\theta_{W} \rho^\nu \left( \check{W}_\mu^- \check{Z}_\nu - \check{W}_\nu^- \check{Z}_\mu \right) \left(\bar{\ell}_{iL} \gamma^\mu \nu_{iL} \right). \;\;\;
\end{eqnarray}
Notice that eq.~\eqref{L3_leptons_lsv} modifies the existing SM vertices, but the operators present in eq.~\eqref{L4_leptons_lsv} are all LSV induced, being initially absent in the SM. With these results, we are ready to consider concrete EW processes.

%%%%%%%%%%%%%%%%%%%%%%%%%%%%%%%%%%%%

\section{Application to selected electroweak processes} \label{sec_app}

\indent

In the previous sections we outlined how LSV may be introduced in the EW sector of the SM. In our more natural scenario, the background 4-vector $\rho$ simultaneously modifies the weak hypercharge and isospin sectors and induces LSV effects wherever $\hat{D}_\mu$ is present. Due to the gauge character of the SM, the first sectors to feel these effects are those of the Higgs and gauge bosons $B_{\mu}$ and $A_{\mu}^{\;\,a} \, (a=1,2,3)$, or rather the mass eigenstates, $A, Z, W^\pm$. Consequently, the interactions of the intermediate bosons with the fermions are also affected. We have, therefore, many options to search for deviations from the expected SM behavior, in particular through high-energy scattering processes.

Before we proceed to analyze actual EW processes, we draw attention to an important practical issue. We have seen that the LSV background effects can be effectively transferred from the quadratic to the interaction terms by means of field redefinitions, cf. eqs.~\eqref{physW} and \eqref{physZ}. This means that the dispersion relations of the mass eigenstates are the same as in the SM, {\it i.e.}, our LSV prescription does not affect the propagation of the vector bosons in vacuum.

In order to calculate cross sections for scattering processes, one needs to consider M\o ller's flux factor $N_1 N_2 |{\bf v}_1 - {\bf v}_2|$, where $N_i$ and ${\bf v}_i$ are the densities and velocities of the incoming particles $i = 1,2$, respectively. This factor quantifies the flux of the colliding beams and is clearly sensitive to the group velocity defined by ${\bf v}_g = \bm{\nabla}_{\bf p} E({\bf p})$, with $E({\bf p})$ coming from the associated dispersion relations. This is an important issue for calculations of decay rates and cross sections in LSV theories, as dully noted in ref.~\cite{Kost_cs}. However, as discussed above, our LSV scenario does not alter the SM dispersion relations for the intermediate gauge bosons or the Higgs -- much less for the fermions. Therefore, the calculations presented below may be performed with the standard tools, since LSV effects will be felt only through the amplitude squared and not flux factors.

%%%%%% Kinematics

\begin{figure}[t!]
\begin{minipage}[b]{0.5\linewidth}
\includegraphics[width=\textwidth]{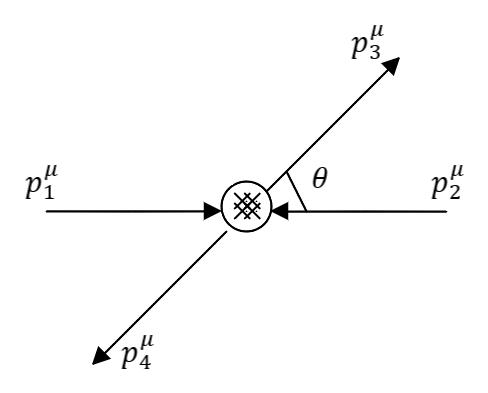}
\end{minipage} \hfill
\caption{Description of the kinematics in the CM frame for the scattering processes considered in this work.}
\label{fig_scattering}
\end{figure}

Finally, let us briefly address the kinematics of the reactions considered here. In the following we will consider three different processes, all of them head-on 2-to-2 scatterings to be analyzed in the center-of-mass (CM) frame, as schematically shown in fig.~\ref{fig_scattering}. Since we will consider processes occurring with a CM energy substantially greater than the electron mass, we may safely neglect it, also in comparison with the masses of the W- and Z-bosons. The incoming particles have 4-momenta given by $p_1^{\;\,\mu} = (E, \mathbf{p})$ and $p_2^{\;\,\mu} = (E, -\mathbf{p})$ and we orient our axes in such a way that the incoming beams lie along the $\hat{z}$ direction, so that $\mathbf{p} = (0, 0, E)$; as mentioned before, here we assume the massless limit, where $E = \vert \mathbf{p} \vert$. For the final momenta, we have two situations to consider, depending if we have equal final states or not. In the first case, we can write $p_3^{\;\,\mu} = (E, \mathbf{q})$ and $p_4^{\;\,\mu} = (E, - \mathbf{q})$, where the final momentum $\mathbf{q}$ is scattered in an angle $\theta$ relative to the initial beam. In the second case, the outgoing particles would still have equal and opposite 3-momenta, but their energies would be different, each of them satisfying the respective on-shell relation $p_i^2 = m_i^2$ and the constraint $E_3 + E_4 = 2E$. Furthermore, for both cases, we are free to define our $\hat{ x}$- and $\hat{y}$-axes such that $\mathbf{q} = \vert \mathbf{q} \vert \, (\sin\theta, 0, \cos\theta)$. At last, but not least, the Mandelstam variables are
\begin{subequations}
\begin{eqnarray}
s & = & \left( p_1 + p_2 \right)^2 = 4E^2 \; , \\
t & = & \left( p_1 - p_3 \right)^2 = m_1^2 + m_3^2 - 2p_1 \cdot p_3 \; , \\
u & = & \left( p_1 - p_4 \right)^2 = m_1^2 + m_4^2 - 2p_1 \cdot p_4 \; .
\end{eqnarray}
\end{subequations}

%%%%%%%%%%%%%%%%%%%%

\subsection{$e^+ \, e^- \rightarrow Z \, H$} \label{sec_eeZH}
\indent

In sec.~\eqref{sec_intro} we briefly discussed how lepton colliders present advantages over hadron colliders when it comes to precise measurements of the SM and possible deviations from it. In this context, let us turn to the process $e^+ \, e^- \rightarrow Z \, H$, also known as Higgs-strahlung. At CM energies around $2 m_H = 250$~GeV, this reaction is by far the most important Higgs production mechanism in $e^+ \, e^-$ colliders, followed by W- and Z-boson fusion~\cite{Kilian, Chen}, which become more important with increasing energy. Higgs-strahlung is a key process to improve our understanding of the Higgs properties in a clean environment: this process allows a precise measurement of the Higgs couplings and also a more complete picture of the Higgs boson decays. This is therefore an ideal process to search for physics beyond the SM, being of central interest in future Higgs factories.

The relevant vertices need to be extracted from the Lagrangians after the field redefinitions. The $ZZH$ vertex may be read directly from eq.~\eqref{L_higgs_lsv}, which already includes the redefined fields; it is given in eq.~\eqref{vert_hzz}. For the $eeZ$ vertex, we use eq.~\eqref{LSMlep} and add it to eq.~\eqref{L3_leptons_lsv}. The result is
\begin{equation}\label{lag_eez}
\mathcal{L}_{eeZ} = \frac{g}{4 \cos\theta_W} \, \bar{\ell}_e \gamma^\mu \left[c_v + c_a \gamma^5\right] \ell_e \left( 1 + \rho \cdot \partial \right) \check{Z}_\mu \;
\end{equation}
and the associated vertex factor is given in eq.~\eqref{vert_eeZ}.

\begin{figure}[t!]
\begin{minipage}[b]{0.5\linewidth}
\includegraphics[width=\textwidth]{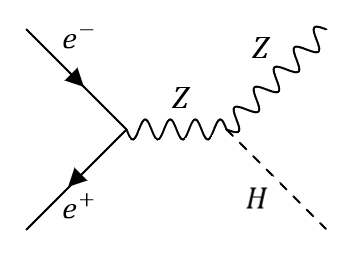}
\end{minipage} \hfill
\caption{The only Feynman diagram contributing to the process $e^-(p_1) \, \, e^+(p_2) \, \, \rightarrow \, \, Z(p_3) \, \, H(p_4)$ at tree-level.}
\label{fig_feyn_eeZH}
\end{figure}

Given the prominent role played by Higgs-strahlung in future precision studies of the EW structure, it is worthwhile determining how our LSV prescription may alter the SM prediction. In fig.~\ref{fig_feyn_eeZH} we show the only Feynman diagram contributing to the amplitude at tree level. Neglecting the electron mass, the unpolarized differential cross section is
\begin{eqnarray} \label{dcs_eeZH}
&& \frac{d \sigma_{eeZH}}{d\Omega} = \frac{G_F^2 m_Z^2}{32 \pi^2} \, (c_v^2 + c_a^2) \, \left(\frac{m_Z^2}{ s - m_Z^2 } \right)^2 \frac{\vert {\bf q} \vert}{\sqrt{s}}    \nonumber \\
& & \times \left[ 1 + \frac{1}{m_Z^2} \left(E_3^2 - \vert {\bf q} \vert^2 \cos^2\theta\right)\right] \, \Pi\left( s, \rho  \right) \; .
\end{eqnarray}
The energy $E_3$ and the module of the 3-momentum $|{\bf q}|$ of the outgoing Z-boson are fixed by kinematics:
\begin{subequations}
\begin{eqnarray}
E_3 & = & \frac{s + m_Z^2 - m_H^2}{2 \sqrt{s}} \; , \label{eq_E3}  \\
\vert {\bf q} \vert & = & \sqrt{  \frac{ \left(s - m_Z^2 - m_H^2\right)^2 - 4 m_Z^2 m_H^2 }{ 4s }} \; .  \label{eq_modq}
\end{eqnarray}
\end{subequations}
The deviation from the SM is controlled by the function
\begin{eqnarray} \label{lsv_factor}
\Pi\left( s, \rho  \right) &=&  1 +  \left[\rho \cdot \left(p_1 + p_2\right) \right]^2 +  \left[\rho \cdot \left(p_1 +  p_2 - p_3\right)  \right]^2 \nonumber \\
&=&  1 +  s\rho_0^2 \left[ 1 + \left( 1 - \frac{ s + m_Z^2 - m_H^2}{2 s}\right)^2  \right]\;,
\end{eqnarray}
where, in the second line, we assume a purely time-like background. The case of a general LSV 4-vector is more involved and is briefly addressed in appendix~\ref{app_B}.

%Notice that, had we considered a purely space-like background, the first LSV term in eq.~\eqref{lsv_factor} would be automatically zero, so that LSV would be felt exclusively in the form of $\Pi\left(s,  {\bm \rho}  \right) = 1 + \left( {\bm \rho}\cdot{\bf q} \right)^2$.

\begin{figure}[t!]
\begin{minipage}[b]{1.\linewidth}
\includegraphics[width=\textwidth]{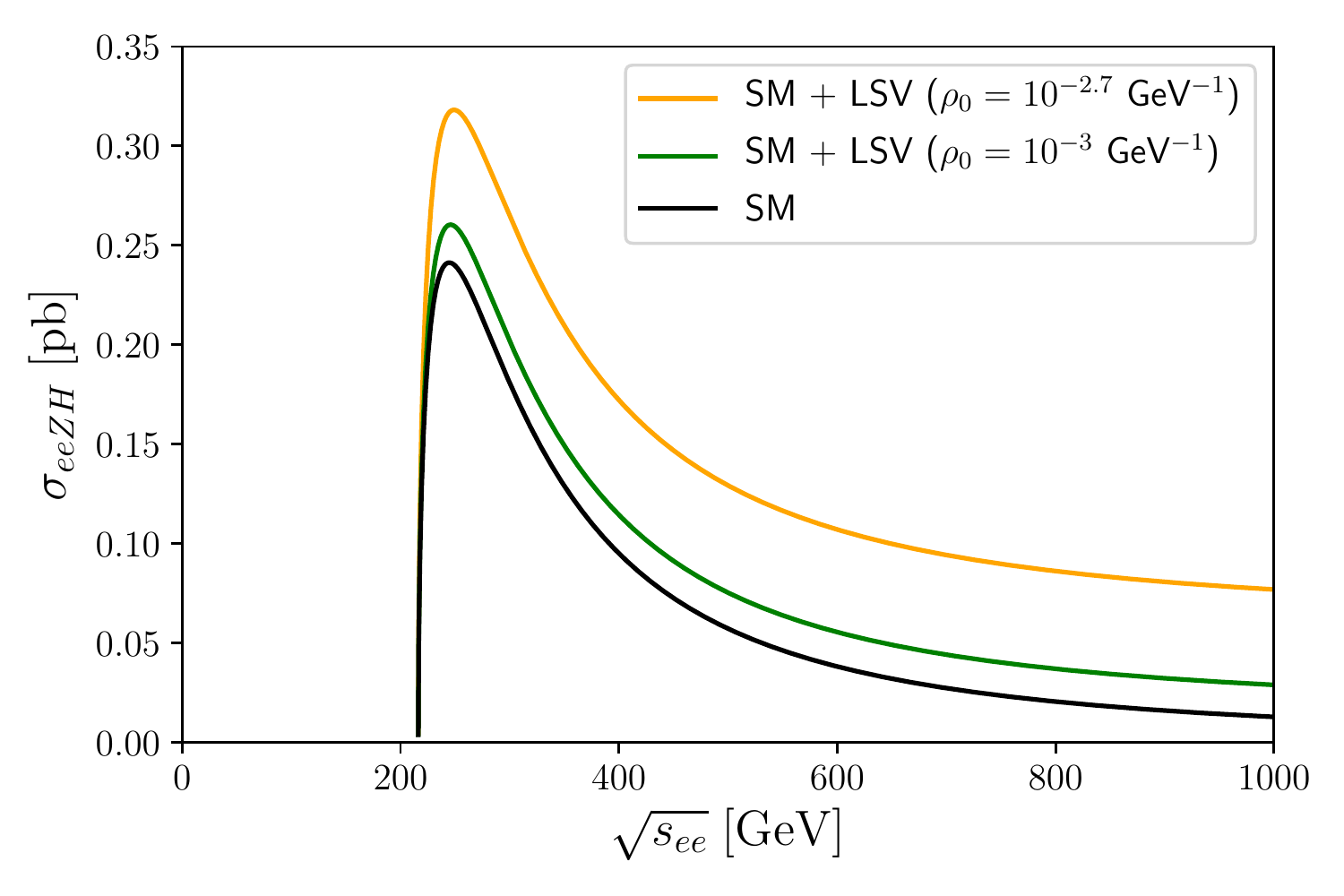}
\end{minipage} \hfill
\caption{Total cross section for $e^+ \, e^- \rightarrow Z \, H $ in the CM. In black is the SM result and in different colors are the combinations of the SM and LSV results (cf.~eq.~\eqref{cs_eeZH}) for various values of the LSV background, here unrealistically large for illustration purposes. }
\label{fig_cs_eeZH}
\end{figure}

Finally, integrating over the solid angle gives us the total cross section
\begin{eqnarray}  \label{cs_eeZH}
\sigma_{eeZH} &=& \frac{G_F^2 m_Z^4 }{96\pi s}  \frac{  (c_v^2 \!  + \!  c_a^2) \lambda^{1/2} \! \left( \lambda \!  + \! 12m_Z^2/s  \right)  }{\left( 1 \!  - \!  m_Z^2/s  \right)^2} \, \Pi\left( s, \rho_0  \right) , \nonumber  \\
\end{eqnarray}
where we defined $\lambda = 4\vert {\bf q} \vert^2/s$ for convenience. The LSV-free part reproduces the SM result, as it should~\cite{Djouadi}. This cross section is shown in fig.~\eqref{fig_cs_eeZH} for different values of the LSV background and for the SM. As expected, the cross section starts from zero at the production threshold $\sqrt{s_{ee}} = m_Z + m_H$, reaches a peak at $\sqrt{s_{ee}} \simeq m_Z + \sqrt{2}m_H$ and then decreases for larger energies. This decay is steeper for the SM case with $\sigma \sim 1/s$, as the LSV factor in $\Pi\left( s, \rho_0  \right)$ compensates this for higher CM energies, thus stabilizing the cross section.

\begin{figure}[t!]
\begin{minipage}[b]{1.\linewidth}
\includegraphics[width=\textwidth]{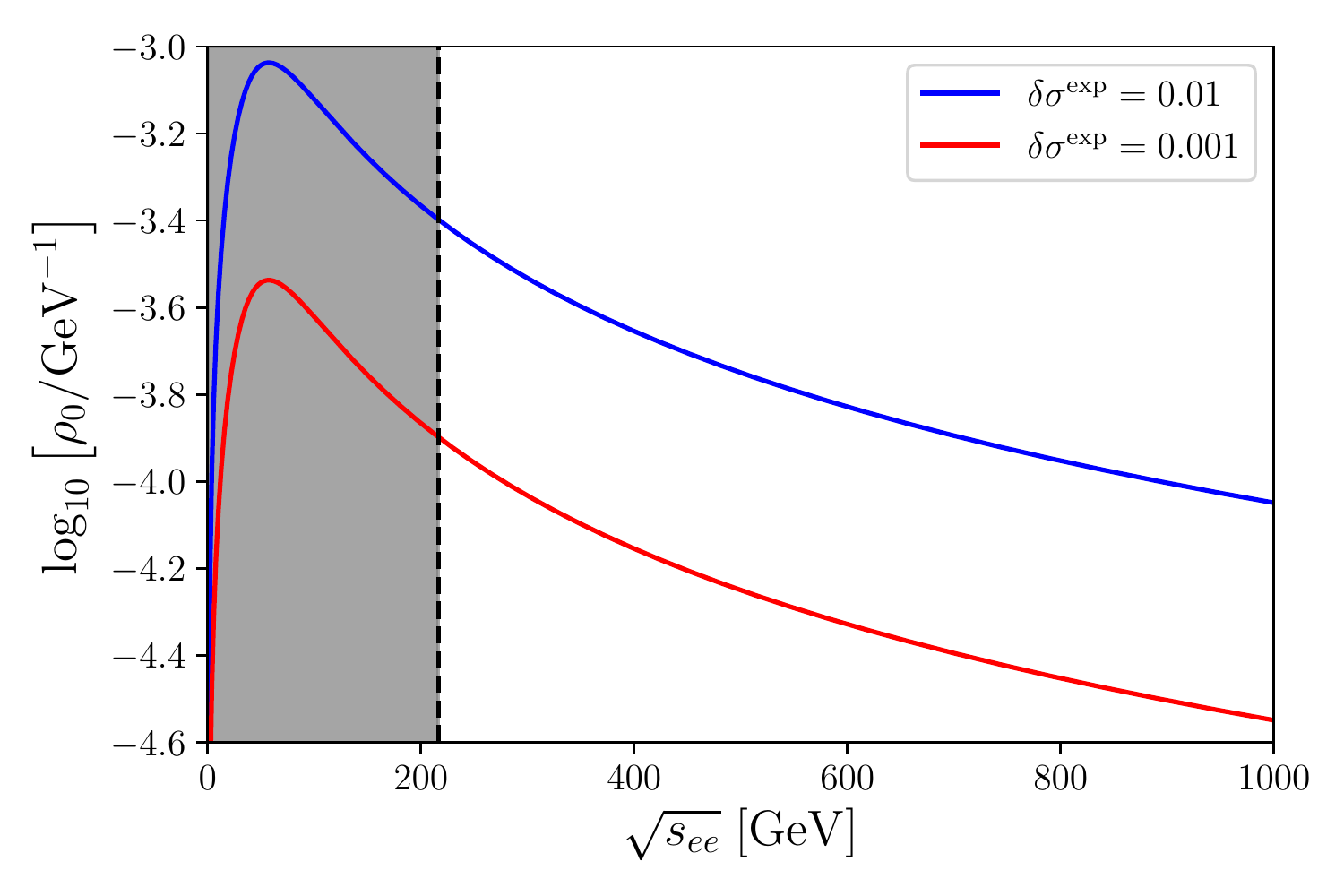}
\end{minipage} \hfill
\caption{Sensitivities for $\rho_0$ for the process $e^+ \, e^- \rightarrow Z \, H$ as a function of the CM energy. The case explicitly discussed in the text is shown in blue, whereas a further ten-fold improvement in the relative uncertainty is shown in red. The vertical dashed line indicates the production threshold $\sqrt{s_{ee}} = m_H + m_Z$ and only the portion of the curves to its right is physically meaningful.}
\label{fig_bound_eeZH}
\end{figure}

Having obtained the LSV-modified cross section, we may discuss possible limits on our time-like background. As stated in ref.~\cite{white}, the ILC will be able to measure $\sigma_{eeZH}$ with $\Delta\sigma/\sigma = \delta\sigma^{\rm exp} \sim 1\%$ at $\sqrt{s_{ee}} = 1$~TeV. If no clear signal of a deviation from the SM prediction is found, we may demand that the LSV effects be hidden within the experimental uncertainties, {\it i.e.},
\begin{equation}   \label{eq_lim_exp_eeZH}
\Bigg| \frac{\sigma^{\rm LSV}_{eeZH}}{\sigma^{\rm SM}_{eeZH}}\Bigg| = \Pi\left( s, \rho_0  \right) - 1  \lesssim \delta\sigma^{\rm exp},
\end{equation}
so that the following projection can be established
\begin{equation} \label{lim_eeZH}
\rho_0 \lesssim 8.9 \times 10^{-5} \, {\rm GeV}^{-1}.
\end{equation}

The behavior of the sensitivities as a function of the CM energy is detailed in fig.~\eqref{fig_bound_eeZH}, also for a ten-fold improvement in the experimental uncertainties. The curves are meaningless for $\sqrt{s_{ee}} \leq m_H + m_Z$, {\it i.e.}, the gray region, since for these CM energies the process itself does not take place, cf. fig.~\eqref{fig_cs_eeZH}. As expected, the sensitivities dramatically improve for high energies. This is, however, not valid for arbitrarily high values, given the non-renormalizable character of our LSV couplings.

%%%%%%%%%%%%%%%%%%%%

\subsection{$e^+ \, e^- \rightarrow Z \, Z$} \label{sec_eeZZ}
\indent

The first evidence for Z-pair production was found in 1999 when LEP-2 reached energies around the threshold for this process~\cite{eezz1,eezz2,eezz3, eezz4}. The study of this channel enables us to check the SM predictions, but it is also sensitive to physics beyond the SM: for instance, it allows us to investigate and constrain anomalous neutral triple gauge couplings $\gamma ZZ$ and $ZZZ$~\cite{eezz5}. Deviations from the SM prediction in this process have been proposed also in the context of two-Higgs models~\cite{eezz6} and low scale gravity~\cite{eezz7}. For Z-pair production in $pp$ collisions, see refs.~\cite{eezz8, eezz9} and references therein.

\begin{figure}[t!]
\begin{minipage}[b]{0.7\linewidth}
\includegraphics[width=\textwidth]{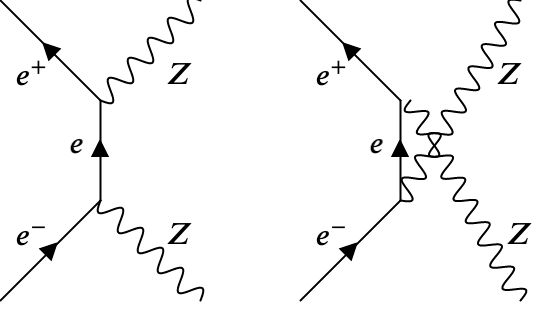}
\end{minipage} \hfill
\caption{The two Feynman diagram contributing to the process $e^-(p_1) \, e^+(p_2) \rightarrow Z(p_3) \, Z(p_4)$ at tree level in the limit of massless electron/positrons ($s \gg m_e^2$).}
\label{fig_feyn_eeZZ}
\end{figure}

At tree level, this process is described by two Feynman diagrams ($t$ and $u$ channels), as depicted in fig.~\ref{fig_feyn_eeZZ}. Note that there is another $s$-channel diagram with a Higgs boson being exchanged. This contribution, however, is determined by the Yukawa couplings, which are proportional to the electron mass. Since we are in energy regimes where $s \gg m_e^2$, we neglect this contribution to the amplitude. The unpolarized differential cross section is given by
\begin{eqnarray}  \label{dcs_eeZZ}
\frac{d\sigma_{eeZZ}}{d\Omega} & = & \frac{G_F^2 m_Z^4 \beta}{256\pi^2 s} \left( c_v^4 +  c_a^4 + 6c_v^2 c_a^2 \right) \nonumber \\
& \times & \frac{A(x) -  \cos^2\theta B(x)  -   \cos^4\theta \beta^4}{\left[ \left( 1 - 2m_Z^2/s \right)^2 - \cos^2\theta \beta^2 \right]^2} \Lambda\left( s, \rho_0  \right) \, ,
\hspace{0.5cm}
\end{eqnarray}
where $\beta = \sqrt{1 - m_Z^2/E^2}$ is the velocity of the outgoing Z-bosons. Setting $x = m_Z/\sqrt{s}$, the coefficients $A(x)$ and $B(x)$ are given by
\begin{subequations}
\begin{eqnarray}
A(x) & = & 1 - 12x^4 + 16x^6 , \\
B(x) & = & 8x^2 - 28x^4 - 16x^6 .
\end{eqnarray}
\end{subequations}

\begin{figure}[t!]
\begin{minipage}[b]{1.\linewidth}
\includegraphics[width=\textwidth]{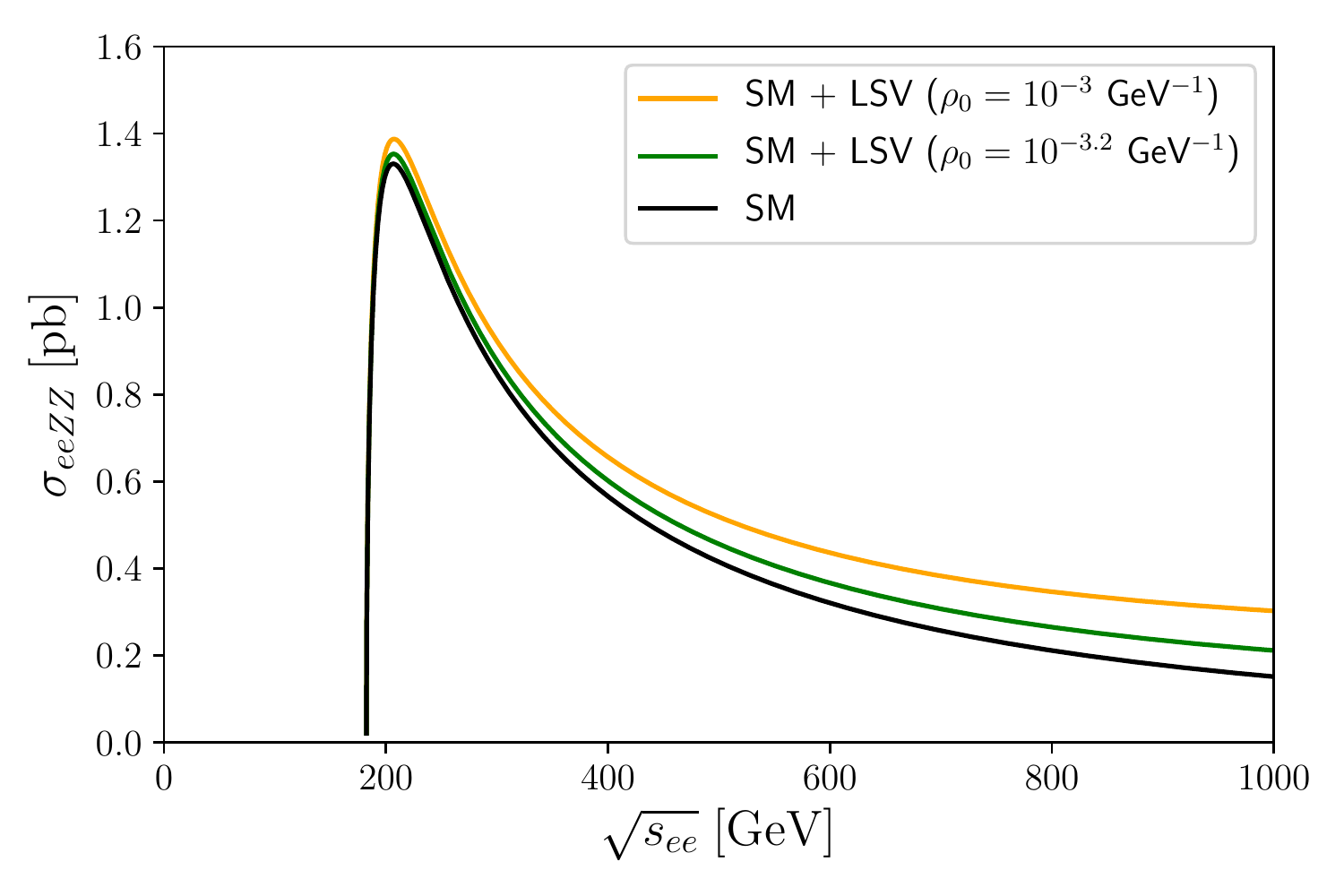}
\end{minipage} \hfill
\caption{Total cross section for $e^+ \, e^- \rightarrow Z \, Z$ in the CM. In black is the SM result and in different colors are the combinations of the SM and LSV results (cf.~eq.~\eqref{cs_eeZZ}) for various values of the LSV background, here unrealistically large for illustration purposes. }
\label{fig_cs_eeZZ}
\end{figure}

The function determining the LSV modification of the SM result turns out to be
\begin{eqnarray} \label{Lambda_LSV}
\Lambda\left(s, \rho  \right) &=& 1 + \left[\rho \cdot \left(p_1+p_2\right)\right]^2 \nonumber \\
&=&  1 + s \rho_0^2 \, .
\end{eqnarray}
It is noteworthy that, in the CM, even for a generic LSV 4-vector, the expression above automatically selects only the time-like component of the background. Notice that the LSV dependence neatly factors out from the SM result, similar to eq.~\eqref{dcs_eeZH}. Equation~\eqref{Lambda_LSV} allows us to easily perform the angular integration and the result is
\begin{eqnarray}  \label{cs_eeZZ}
\sigma_{eeZZ} & = & \frac{G_F^2 m_Z^4 }{32\pi s} \left( c_v^4 +  c_a^4 + 6c_v^2 c_a^2 \right) \Lambda\left( s, \rho_0  \right) \nonumber \\
& \times & \left[  \frac{1 + 4m_Z^4/s^2}{1 - 2m_Z^2/s} \log\left(\frac{1 + \beta}{1 - \beta} \right)  - \beta  \right] ,
\end{eqnarray}
where $c_{v,a}$ were defined in eq.~\eqref{c_va}. This result is shown in fig.~\eqref{fig_cs_eeZZ} and, as required, the LSV-free part matches the tree-level SM result~\cite{rwbrown}.

Let us now briefly discuss possible upper bounds on the time-like LSV background. As usual, we simply compare the relative experimental uncertainty with the expected deviation from the SM by analyzing
\begin{equation}
\Bigg| \frac{\sigma^{\rm LSV}_{eeZZ}}{\sigma^{\rm SM}_{eeZZ}}\Bigg| = \Lambda\left( s, \rho_0  \right) - 1  \lesssim \delta\sigma^{\rm exp},
\end{equation}
from which we may extract an upper limit on the LSV parameter.

The process $e^+ \, e^- \rightarrow Z \, Z$ is not easily measured, as the final products are short lived. Furthermore, the CM energy required to even make this reaction possible, $\sqrt{s_{ee}} = 2m_Z = 182.4$~GeV, was inaccessible a couple of decades ago. For instance, the ALEPH and OPAL collaborations analyzed data from LEP at CM energies around 200~GeV~\cite{eezz2, eezz5}. Their results match the SM prediction, but carry $\sim 20\%$ uncertainties, thus leading to the overly weak projection $\rho_0 \lesssim 2.4 \times 10^{-3} \, {\rm GeV}^{-1}$. Since this result is more than 20 years old, we expect that future, state-of-the-art $e^+ \, e^-$ colliders will be able to achieve $\delta\sigma^{\rm exp} \sim 1\%$ at TeV energies. Under these conditions we could constrain the time-like LSV background to be
\begin{equation} \label{lim_eeZZ}
\rho_0 \lesssim 1.0 \times 10^{-4} \, {\rm GeV}^{-1}.
\end{equation}

\begin{figure}[t!]
\begin{minipage}[b]{1.\linewidth}
\includegraphics[width=\textwidth]{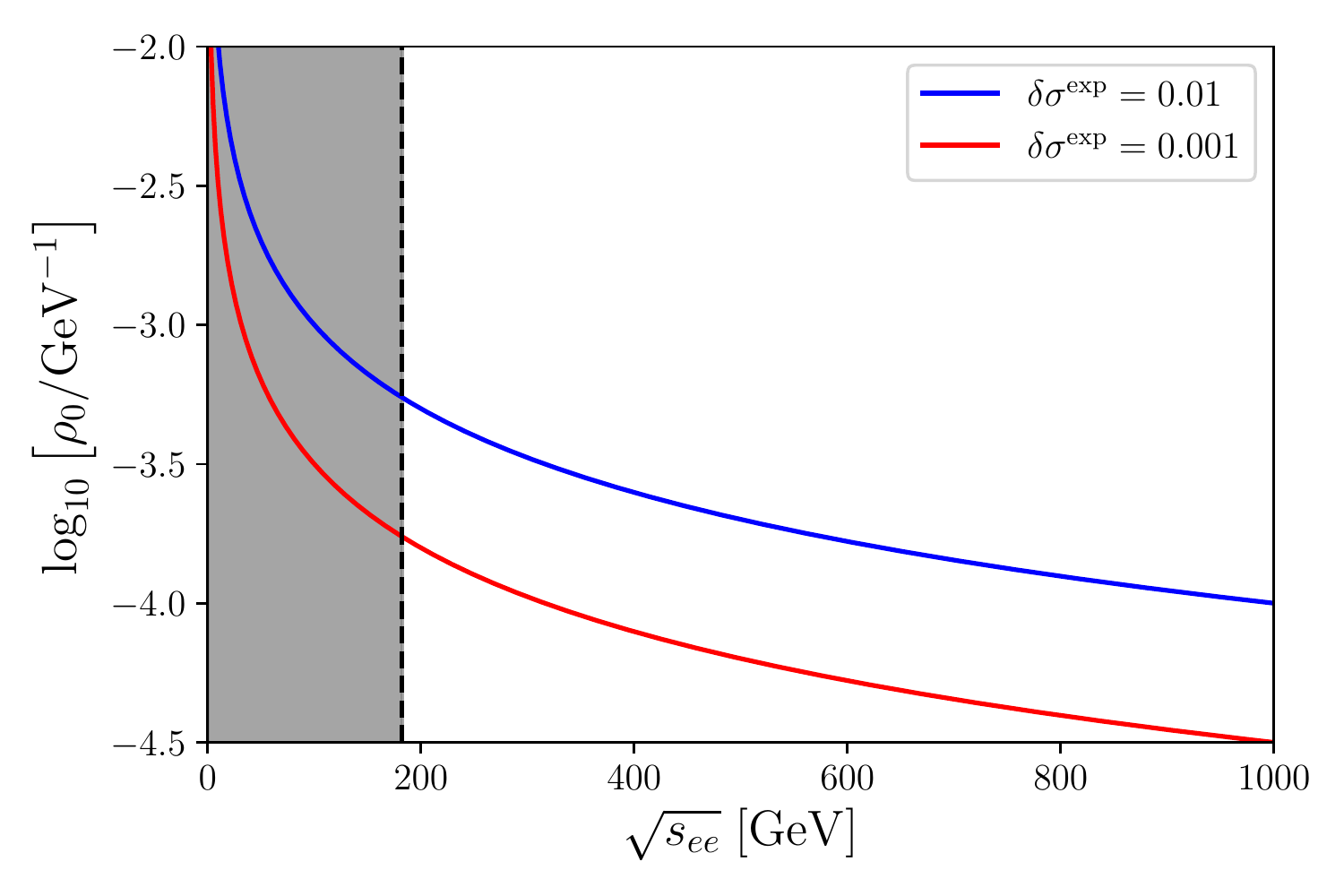}
\end{minipage} \hfill
\caption{Sensitivities for $\rho_0$ for the process $e^+ \, e^- \rightarrow Z \, Z$ as a function of the CM energy. The case explicitly discussed in the text is shown in blue, whereas a further ten-fold improvement in the relative uncertainty is shown in red. The vertical dashed line indicates the production threshold $\sqrt{s_{ee}} = 2m_Z$ and only values to its right are physically meaningful.}
\label{fig_bound_eeZZ}
\end{figure}

The general behavior of the bounds for different relative uncertainties is illustrated in fig.~\eqref{fig_bound_eeZZ} in which, similar to the case of the previous section, the only meaningful values are those above threshold.

%%%%%%%%%%%%%%%%%%%%

\subsection{$\gamma \, \gamma \rightarrow W^+ \, W^-$} \label{sec_AAWW}

\indent

%The SM is based on a gauge symmetry, a fact manifest in the necessity of a covariant derivative, cf.~eq.~\eqref{cov_dev}. Furthermore, the gauge group of the SM is non-Abelian, what leads, among other things, to non-linearities in the field-strength tensors. This is important because the kinetic terms produce not only quadratic, but also triple and quartic (self-interaction) couplings, something unique to non-Abelian theories. In fact, the measurement of processes involving such couplings is an important test of the structure of the SM~\cite{Yehudai, Tupper, Choi, Eboli, Ari}.

\begin{figure}[t!]
\begin{minipage}[b]{1.\linewidth}
\includegraphics[width=\textwidth]{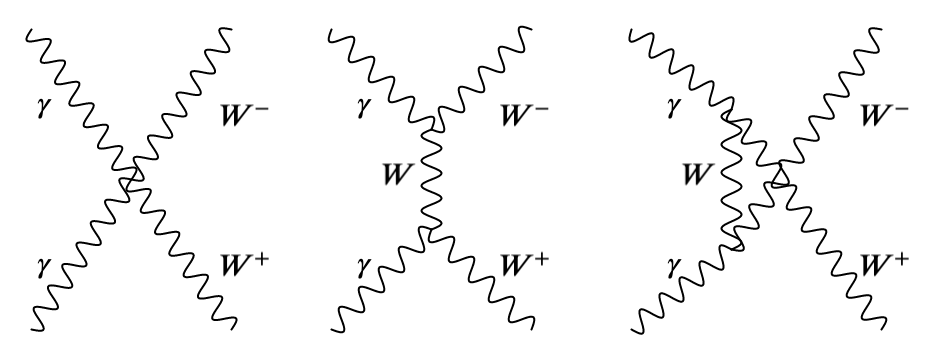}
\end{minipage} \hfill
\caption{Feynman diagrams contributing to the process $\gamma(p_1) \, \gamma(p_2) \rightarrow W^-(p_3) \, W^+(p_4)$ at tree level.}
\label{fig_feyn_AAWW}
\end{figure}

The measurement of processes involving gauge couplings is an important test of the structure of the SM~\cite{Yehudai, Tupper, Choi, Eboli, Ari}. In particular, the vertices $\gamma \, W^+ \, W^-$ and $\gamma \, \gamma \, W^+ \, W^-$ allow for the photon-photon production of a pair of W-bosons via three Feynman diagrams already at tree level, cf. fig.~\ref{fig_feyn_AAWW}. The tree-level unpolarized cross section in the CM is given by~\cite{Denner95, Denner96}
\begin{eqnarray} \label{dcs_AAWW_SM}
\frac{d\sigma^{\rm SM}_{\gamma \gamma W W}}{d\Omega} & = & \frac{3\alpha^2 \beta}{2s} \Bigg[ 1 - \frac{2s\left( 2s + 3m_W^2  \right)}{3\left( m_W^2 - t  \right) \left( m_W^2 - u  \right)} \nonumber \\
& + & \frac{2s^2\left( s^2 + 3m_W^4  \right)}{3\left( m_W^2 - t  \right)^2 \left( m_W^2 - u  \right)^2} \Bigg] \; ,
\end{eqnarray}
where $\alpha = e^2/4\pi$ is the electromagnetic fine-structure constant. The Mandelstam variables, expressed in terms of the energy $E$ of the incoming photon and the scattering angle $\theta$, are
\begin{subequations}
\begin{eqnarray}\label{mandelstam}
s & = & 4E^2 \; , \\
t & = & m_W^2 - 2E^2\left(1 - \beta\cos\theta \right) \; , \\
u & = & m_W^2 - 2E^2\left(1 + \beta\cos\theta \right) \; .
\end{eqnarray}
\end{subequations}
Here $\beta = \sqrt{1 - m_W^2/E^2}$ is the velocity of the outgoing W-bosons. The total SM differential cross section integrated within the finite angular acceptance of the detector ($\theta_{\rm cut} \leq \theta \leq \pi - \theta_{\rm cut}$) is~\cite{Denner95, Denner96}
\begin{eqnarray} \label{cs_AAWW_SM}
\sigma^{\rm SM}_{\gamma \gamma W W} & = & \frac{6\pi\alpha^2}{s} \beta\cos\theta_{\rm cut} \bigg[ 1 + \frac{16\left( s^2 + 3m_W^4 \right)}{3s^2 \left( 1 - \beta^2 \cos^2 \theta_{\rm cut}  \right)} \nonumber \\
& - & \! \frac{4m_W^2 \left( s - 2m_W^2 \right)}{s^2 \beta \cos \theta_{\rm cut} } \log\left(\frac{1 + \beta \cos \theta_{\rm cut}}{1 - \beta \cos \theta_{\rm cut}} \right) \! \bigg].
\hspace{0.5cm}
\end{eqnarray}
%
%The reason for keeping a finite cut-off angle will become clear shortly.

We may now look at our LSV scenario at leading order. After redefining the W- and Z-fields, we find that the coupling $\gamma \, W^{+} W^{-}$ is governed by
\begin{eqnarray}\label{L_AWW}
\mathcal{L}_{\gamma WW} & = & i e \left[ F^{\mu \nu} \check{W}_\mu^+ \check{W}_\nu^- - A^\mu \left( \check{W}_{\mu \nu}^+  \check{W}^{\nu -} - \check{W}_{\mu \nu}^-  \check{W}^{\nu +} \right) \right]  \nonumber \\
& + & i e \rho^\sigma F^{\mu \nu} \left( \check{W}_\mu^+ \partial_\nu \check{W}_\sigma^- + \check{W}_\nu^- \partial_\mu \check{W}_\sigma^+ \right)  \nonumber \\
& - & i e  \rho^\sigma A^{\mu} \left( \check{W}_{\mu\nu}^+ \partial^\nu \check{W}_\sigma^- - \check{W}_{\mu\nu}^- \partial^\nu \check{W}_\sigma^+  \right)   \nonumber \\
& - & i e m_W^2 \rho^\nu A^\mu  (\check{W}_\mu^+ \check{W}_\nu^- - \check{W}_\mu^- \check{W}_\nu^+) \; ,
\end{eqnarray}
whereas for the coupling $\gamma \gamma \, W^{+} W^{-}$ we have
\begin{eqnarray}\label{L_AAWW}
\mathcal{L}_{\gamma\gamma WW} & = & e^2 \left( A^\mu A^\nu \check{W}_\mu^+  \check{W}_\nu^- - A_\mu A^\mu \check{W}_\nu^+ \check{W}^{\nu -} \right) \nonumber \\
& + & e^2 \rho^\sigma A^\alpha \bigg[  A^\beta \left( \check{W}_\beta^- \partial_\alpha \check{W}_\sigma^+ + \check{W}_\alpha^+ \partial_\beta \check{W}_\sigma^-  \right)
\nonumber \\
& - & A_\alpha  \left( \check{W}_\beta^- \partial_\beta \check{W}_\sigma^+ + \check{W}_\beta^+ \partial_\beta \check{W}_\sigma^-  \right)  \bigg] \; . \;
\end{eqnarray}
The respective Feynman rules may be readily obtained and are conveniently stated in eqs.~\eqref{vert_AWW} and~\eqref{vert_AAWW}.

The differential cross section in the CM shows once again no first-order term in the LSV 4-vector. Specializing to a purely time-like background, the differential cross section is the sum of the SM contribution, cf. eq.~\eqref{cs_AAWW_SM}, and the second-order LSV contribution given by
\begin{eqnarray} \label{dcs_AAWW_LSV}
&&\frac{d\sigma^{\rm LSV}_{\gamma \gamma W W}}{d\Omega} = \frac{\alpha^2}{8192} \frac{s^3}{m_W^6} \frac{\beta^3 \rho_0^2}{\beta^2 \cos^2\theta - 1} \nonumber \\
&& \times \left[ \, C(x) -  4 \cos(2\theta) \, D(x) + \cos(4\theta) \, E(x) \, \right] \; , \;
\end{eqnarray}
where we defined
\begin{subequations}
\begin{eqnarray}
C(x) & \!=\! &  11 + 156x^2 - 672x^4 + 4352x^6 + 6144x^8 \, , \hspace{0.2cm}
\\
D(x) & \!=\! & 3 - 12x^2  - 192x^4 - 320x^6 + 512x^8 \; ,   \\
E(x) & \!=\! &  1 - 12x^2 + 32x^4 \; ,
\end{eqnarray}
\end{subequations}
with $x =m_W/\sqrt{s}$. Finally, the total LSV differential cross section integrated in $\theta_{\rm cut} \leq \theta \leq \pi - \theta_{\rm cut}$ is
\begin{eqnarray} \label{cs_AAWW_LSV}
\sigma^{\rm LSV}_{\gamma \gamma W W} & = & -\frac{\pi\alpha^2\beta\cos\theta_{\rm cut}}{256} \frac{s^3 \rho_0^2}{m_W^6} \bigg\{ \! \tilde{C}(x) - \frac{\beta^2 \cos^2\theta_{\rm cut} }{3} \tilde{D}(x)
\nonumber \\
& + &  \frac{\tilde{E}(x)}{\beta\cos\theta_{\rm cut}}  \log\left(\frac{1 + \beta \cos \theta_{\rm cut}}{1 - \beta \cos \theta_{\rm cut}} \right)  \! \bigg\} ,
\end{eqnarray}
with
\begin{subequations}
\begin{eqnarray}
\tilde{C}(x) & = & 3 - 16x^2 \left( 1 + 10x^2 + 20x^4 - 32x^6 \right) \; ,
\\
\tilde{D}(x) & = & 1 - 8x^2  \; ,
\\
\tilde{E}(x) & = & 8x^2 \left( 1 - 4x^2 + 88x^4 - 64x^6 - 256x^8 \right).
\hspace{0.3cm}
\end{eqnarray}
\end{subequations}

\begin{figure}[t!]
\begin{minipage}[b]{1.\linewidth}
\includegraphics[width=\textwidth]{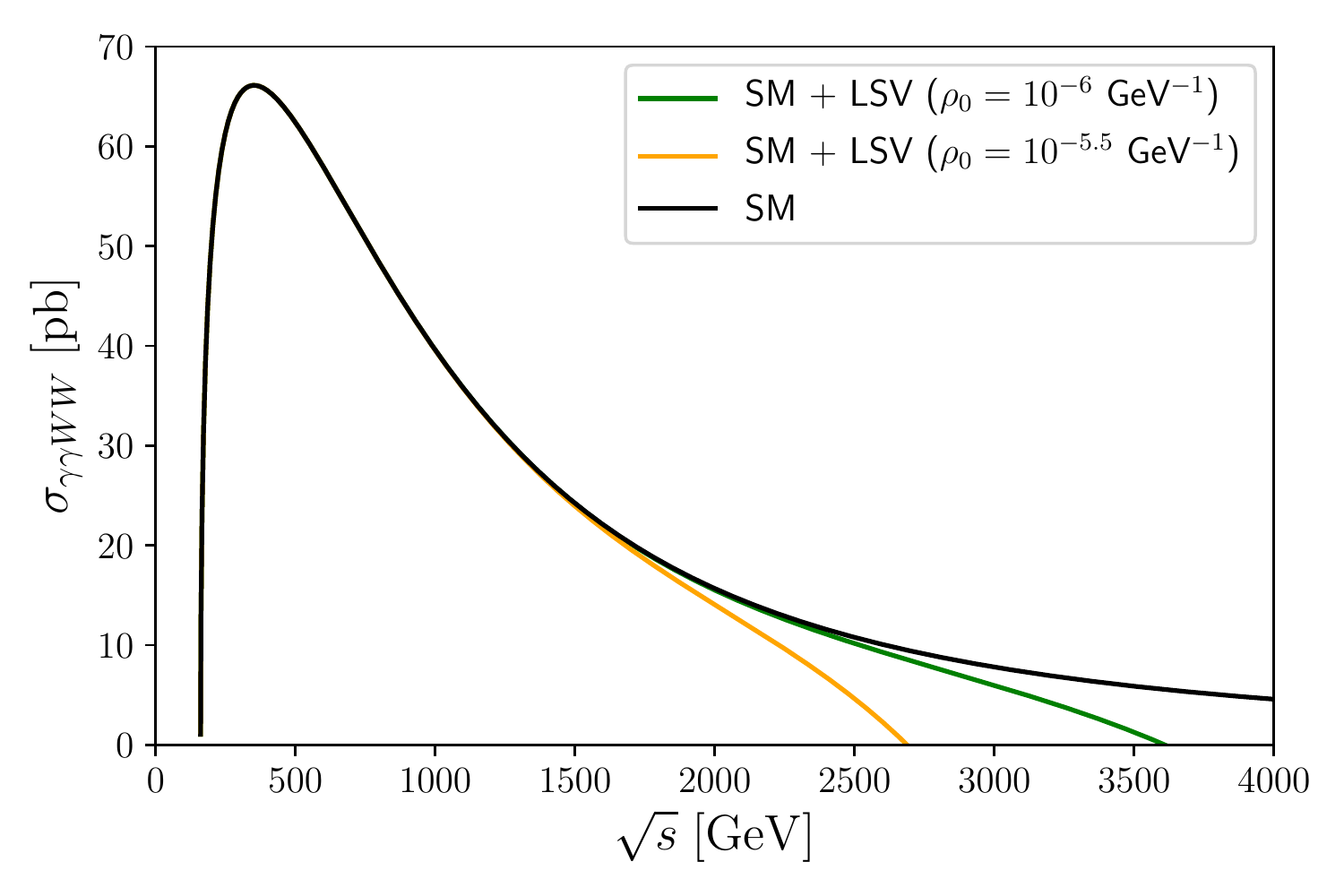}
\end{minipage} \hfill
\caption{Total cross sections for $\gamma \gamma \rightarrow W W$ in the CM with $\theta_{\rm cut} = 9.4^\circ$. In black is the SM result (cf.~eq.~\eqref{cs_AAWW_SM}) and in different colors are the combinations of the SM and LSV results (cf.~eq.~\eqref{cs_AAWW_LSV}) for various values of the LSV background. As expected, the cross section starts from zero at the production threshold $\sqrt{s_{\gamma\gamma}} = 2 m_W$, reaches a peak and then monotonically decreases for larger energies.}
\label{fig_cs_AAWW}
\end{figure}

The total cross section is shown in fig.~\eqref{fig_cs_AAWW} for different values of the time-like LSV background. Contrary to the other cases considered previously, the leading-order LSV contribution is negative, thus reducing the overall cross section. At this point we are able to justify the introduction of the cut-off angle. Equation~\eqref{cs_AAWW_LSV} is such that the total cross section cannot be cast in the form $\sigma_{\gamma \gamma W W} = \left( 1 + \rho_0^2 \cdot f(s)  \right) \sigma^{\rm SM}_{\gamma \gamma W W}$, with $f(s)$ being some function only of the CM energy, as in eqs.~\eqref{cs_eeZH} and~\eqref{cs_eeZZ}. This is a consequence of the more involved momentum structure of the $\gamma WW$ and $\gamma\gamma WW$ vertices, cf. eqs.~\eqref{vert_AWW} and~\eqref{vert_AAWW}, which introduces extra angular factors in the squared amplitude and hence in the differential cross section.

Measuring the process $\gamma \gamma \rightarrow W^{+} W^{-}$ is important in future experiments, as it provides a valuable possibility of testing anomalous gauge couplings. The photons in this process are usually produced in one of two ways: {\it bremsstrahlung}~\cite{Yellin} or Compton back scattering~\cite{Kim}. The former is typical in standard colliders, such as LEP and LHC, but has the disadvantage that the photons are soft with a luminosity decreasing for larger energies. The latter, on the other hand, is very interesting, since the photons end up with around $80\%$ of the initial CM energy, {\it i.e.}, $\sqrt{s_{\gamma\gamma}} \approx 0.8 \sqrt{s}$~\cite{Telnov}. For the sake of concreteness, let us consider the energies envisaged for linear $e^+ e^-$ colliders, such as the TESLA~\cite{tesla} or the ILC~\cite{ilc, ilc2}, which would use the Compton back scattering of electrons in powerful lasers to set up $\gamma\gamma$ or $e\gamma$ collisions. The CM energies involved would reach as much as $\sqrt{s} = 500$~GeV (TESLA) and $\sqrt{s} = 1000$~GeV (ILC), so that the available energy for the colliding photons would be $\sqrt{s_{\gamma\gamma}} \approx 400$~GeV and $\sqrt{s_{\gamma\gamma}} \approx 800$~GeV, respectively.

\begin{figure}[t!]
\begin{minipage}[b]{1.\linewidth}
\includegraphics[width=\textwidth]{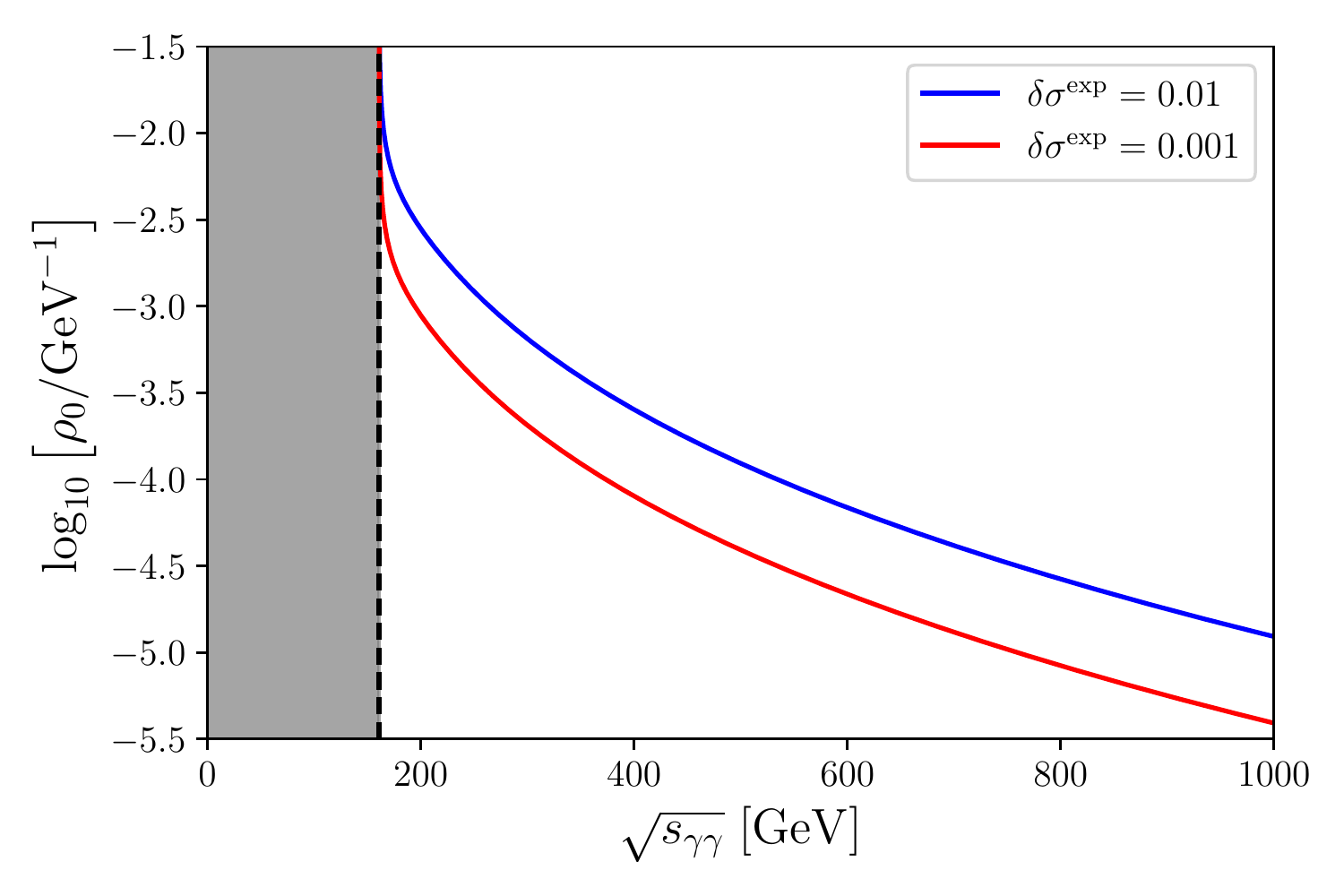}
\end{minipage} \hfill
\caption{Sensitivities for $\rho_0$ for the process $\gamma \gamma \rightarrow W^{+} W^{-}$ as a function of the CM energy (using $\theta_{\rm cut} \simeq 9.4^\circ$). The case explicitly discussed in the text is shown in blue, whereas a further ten-fold improvement in the relative uncertainty is shown in red. The vertical dashed line indicates the production threshold $\sqrt{s_{\gamma\gamma}} = 2 m_W$ and only values to its right are physically meaningful.}
\label{fig_bound_AAWW}
\end{figure}

The process $\gamma \gamma \rightarrow W^{+} W^{-}$ has not yet been directly measured, but it does take place, albeit rarely, at the LHC in peripheral pp collisions via $pp(\gamma\gamma) \rightarrow p^{(*)} WW p^{(*)}$. Working with data taken at $\sqrt{s} = 13$~TeV, the measured cross section was $\sigma^{\rm exp} = 3.13 \pm 0.42$~fb (statistical and systematic uncertainties added in quadrature), which is in agreement with the SM prediction~\cite{LHC_WW}. In order to estimate the LSV background, we assume that future photon colliders would reach an optimistic ten-fold better relative uncertainty than the $\sim 13\%$ quoted above, {\it i.e.}, reach $\delta\sigma^{\rm exp} = 1\%$. Considering that the SM predictions for $\gamma \gamma \rightarrow W^{+} W^{-}$ still match the data, we may obtain an upper bound on the time-like LSV background by demanding that
\begin{equation}
\Bigg| \frac{\sigma^{\rm LSV}_{\gamma \gamma W W}}{\sigma^{\rm SM}_{\gamma \gamma W W}}\Bigg|  \lesssim \delta\sigma^{\rm exp} \, .
\end{equation}

Assuming the $\sqrt{s_{\gamma\gamma}} = 800$~GeV attainable at the ILC and an angular acceptance similar to that of the ATLAS detector~\cite{LHC_WW}, {\it i.e.}, a cut-off angle $\theta_{\rm cut} \simeq 9.4^\circ$ (corresponding to a pseudo-rapidity $|\eta| < 2.5$), we get
\begin{equation} \label{lim_AAWW}
\rho_0 \lesssim 2.7 \times 10^{-5} \, {\rm GeV}^{-1} \, .
\end{equation}
In fig.~\eqref{fig_bound_AAWW} we plot the sensitivity as a function of the available CM energy. The steepness of the energy dependence is due to the fact that the LSV cross section grows as $| \sigma^{\rm LSV}_{\gamma \gamma W W} | \sim s^3$ at high energies, cf. eq.~\eqref{cs_AAWW_LSV}.

%%%%%%%%%%%%%%%%%%%%%%%%
	
\section{Concluding remarks} \label{sec_conclusion}
\indent

In this paper we investigate the processes $e^{+} \, e^{-} \rightarrow Z \, H$, $e^{+} \, e^{-} \rightarrow Z \, Z$ and $\gamma \, \gamma \rightarrow W^{+}W^{-}$ at tree level assuming that the SM is modified by LSV non-minimal couplings in the EW sector as given by eq.~\eqref{cov_dev_lsv}. We obtain the field equations for the gauge bosons and, focusing on the case $\rho = \xi$, we find that the usual dispersion relations may be obtained by employing a convenient field redefinition, rendering the kinetic terms LSV-free and effectively transferring the LSV dependence to the interaction terms. We then write the relevant Lagrangians for the Z- and W-bosons, as well as for the Higgs scalar and the leptons, extracting the interaction vertices and calculating the unpolarized cross sections of the aforementioned processes, cf. sec.~\ref{sec_app}. We note that the leading-order LSV contributions are of $\mathcal{O}\left( \rho^2 \right)$, a fact also encountered in refs.~\cite{Mario, Gomes, scat_LSV, mouchrek1, mouchrek2}.

From eqs.~\eqref{cs_eeZH}, \eqref{cs_eeZZ}, \eqref{cs_AAWW_SM} and~\eqref{cs_AAWW_LSV} we obtain sensitivities for the case of a purely time-like LSV background. All the LSV-modified cross sections display a somewhat similar behavior: the SM contributions typically decay as $1/s$, but the respective LSV corrections counteract this tendency for higher energies. This means that the bounds would be harder for higher CM energies, an expected conclusion, since we are dealing with effective, non-renormalizable couplings that involve the field-strength tensors carrying an extra factor of 4-momentum into the vertices.

Currently, the SM predictions are broadly confirmed experimentally. This implies that new physics, if any, must be buried under the experimental uncertainties. Next-generation $e^+ e^-$ colliders will probe the SM with unprecedented precision and provide an outstanding opportunity to search for possible deviations. Ideally, we would find clear signals of new physics, but, assuming that the SM is further confirmed, we are able to establish upper limits -- or rather, projections -- on the time-like component of the LSV background. Our strongest bound comes from photon-induced W-pair production and is found to be of the order
\begin{equation}
\rho_0^{\rm SCF} \lesssim 10^{-5} \, {\rm GeV}^{-1} \; ,
\end{equation}
which is already expressed in the SCF, as discussed in sec.~\ref{sec_intro}. This result is compatible with the bounds obtained elsewhere~\cite{Mario}. We note that, due to the quadratic dependence of the LSV cross sections on $\rho_0$, the bounds improve only with $\sim \sqrt{\delta\sigma^{\rm exp}}$, so that small increments in experimental precision would result in only marginally better limits.

So far we have focused exclusively on a purely time-like LSV background, so let us briefly comment on the space-like case. First of all, as highlighted in sec.~\ref{sec_intro}, the space components in the laboratory are related to those in the SCF by a time-dependent matrix $R(\chi,T)$ associated with Earth's motion~\cite{ref_sun}. Not all processes in our analysis would behave the same in the presence of a space-like background, though. For instance, in $e^-e^+\rightarrow Z Z$, a purely LSV background would not give any correction to the SM cross section, cf. eq.~\eqref{Lambda_LSV}. On the other hand, in the case of $e^-e^+ \rightarrow Z H$, we see that a purely space-like LSV background would give a correction of the form $\sim \left( \vert \bm{\rho} \vert \vert\textbf{q}\vert \cos\zeta \right)^2 $, where $\textbf{q}$ is the momentum of the Z-boson and $\zeta$ is the angle between ${\bm \rho}$ and $\textbf{q}$, cf. eq.~\eqref{lsv_factor}. This angle may be expressed in terms of the polar and azimuthal angles in the usual coordinate system of the incoming beams, so that the cross section would not only develop a time dependence as the background appears to rotate, but also acquire an atypical dependence on the azimuthal angle, cf. refs.~\cite{scat_LSV, Gomes}. A similar analysis of the process $\gamma \gamma \rightarrow W^{+} W^{-}$ would be far more complicated, but the conclusion would be analogous. For a discussion of similar scenarios in the context of QED we refer the reader to refs.~\cite{scat_LSV, Gomes}.

To conclude our contribution, let us say a few words about possible future developments. Firstly, it would be interesting to analyze scattering processes considering a purely spatial background in different directions. Another possibility would be to work out the consequences of having different LSV backgrounds associated with each gauge group, {\it i.e.}, $\rho \neq \xi$. As we already remarked, in this scenario there is a $\gamma-Z$ mixing that could have potentially interesting phenomenological implications. Moreover, in this case we would also have an anomalous $\gamma H Z$ vertex that would induce -- at tree level -- the decay $H\rightarrow\gamma Z$, which is a loop-mediated process in the SM~\cite{HAZ1, HAZ2}.

Also, given the energy scales considered here, one could readily extend our LSV prescription to include the quark sector and study for instance LSV effects in top-quark pair production~\cite{top}. Finally, in the light of the recent measurements of the muon $(g-2)$ at Fermilab~\cite{muon}, one could expand our prescription to allow family-dependent LSV parameters and determine if isotropic LSV backgrounds could explain the reported $4.2 \, \sigma$ deviation from the SM prediction.

%Finally, one could slightly modify our prescription to allow flavor-dependent LSV parameters, and we wonder if it would be possible to adjust these parameters in such a way that one could consistently explain the recently measured anomalous magnetic dipole moment of the muon~\cite{muon}.

%%%%%%%%%%%%%%%%%%%%%%%%%%%%%%%%
%	
\begin{acknowledgments}
The authors are grateful to J. A. Helay\"el-Neto for reading the manuscript and useful comments. We are also grateful to the anonymous referee for his/her helpful suggestions. MJN thanks CNPq (Conselho Nacional de Desenvolvimento Cient\' ifico e Tecnol\'ogico), the Brazilian scientific support federal agency, for partial financial support, grant number 313467/2018-8. PDF thanks G.P. de Brito for interesting discussions and CNPq for funding.
\end{acknowledgments}
%	

%%%%%%%%%%%%%%%%
\appendix
	
\section{Feynman rules for LSV vertices} \label{app_A}
\indent

In the main text we have discussed how convenient field redefinitions in the gauge sector move the LSV background from the quadratic part of the equations of motion to the interaction terms. Though straightforward, these redefinitions produce cumbersome equations which we only presented explicitly for the processes considered in sec.~\ref{sec_app}. Here we collect the Feynman rules resulting from the Lagrangians expressed in terms of the redefined fields, $\check{W}$ and $\check{Z}$.

The vertices needed to compute the cross sections in secs.~\ref{sec_eeZH} and~\ref{sec_eeZZ} are given by:

%\feynmandiagram [baseline=(a.base), horizontal=a to b] {
%	i1 [particle=\(Z_\mu\)] -- [boson, momentum=\(p_{1}\)] a -- [boson, rmomentum=\(p_{2}\)] i2 [particle=\(Z_{\nu}\)],
%	a -- [scalar] b [particle=\(h\)],
%};
%\feynmandiagram [baseline=(a.base), horizontal=a to b] {
%	i1 [particle=\(e^-\)] -- [fermion, momentum=\(p_{1}\)] a -- [fermion, rmomentum=\(p_{2}\)] i2 [particle=\(e^+\)],
%	a -- [boson] b [particle=\(Z_\mu\)],
%};
\vspace{0.3cm}
\begin{tikzpicture}
	\begin{feynman}
		\vertex (a) {\(H\)};
		\vertex [left=of a] (b);
		\vertex [above left=of b] (f1) {\(Z_\mu\)};
		\vertex [below left=of b] (c){\(Z_\nu\)};
		\diagram* {
			(a) -- [scalar] (b) -- [boson, rmomentum'=\(p_1\)] (f1),
			(b) -- [boson, rmomentum=\(p_2\)] (c)		};
	\end{feynman}
\end{tikzpicture}
\begin{tikzpicture}
	\begin{feynman}
		\vertex (a) {\(Z_\mu\)};
		\vertex [left=of a] (b);
		\vertex [above left=of b] (f1) {\(e^-\)};
		\vertex [below left=of b] (c){\(e^+\)};
		\diagram* {
			(a) -- [boson] (b) -- [anti fermion, rmomentum'=\(p_1\)] (f1),
			(b) -- [fermion, rmomentum=\(p_2\)] (c)		};
	\end{feynman}
\end{tikzpicture}

\noindent
and the respective vertex factors are
\begin{eqnarray}
\mathcal{V}_{\mu\nu}^{(HZZ)}  &=& 2  i \, \frac{m_z^2}{ v} \, \eta_{\mu \nu} [1 - i \rho \cdot (p_1 + p_2)] \; ,
\label{vert_hzz} \\
\mathcal{V}^{(eeZ)}_{\mu} &=& i \, \frac{m_z}{2 v} \, \gamma_\mu (c_v + c_a \gamma^5) [1 + i \rho \cdot (p_1 + p_2)] \, .
\hspace{0.6cm}
\label{vert_eeZ}
\end{eqnarray}

The triple and quartic vertices used for sec.~\ref{sec_AAWW} and the respective vertex factors are:
\vspace{0.42cm}
\begin{tikzpicture}
	\begin{feynman}
		\vertex (a) {\(A_\lambda\)};
		\vertex [right=of a] (b);
		\vertex [above right=of b] (f1) {\(W_\mu^-\)};
		\vertex [below right=of b] (c){\(W_\nu^+\)};
		\diagram* {
			(a) -- [boson, momentum=\(p_3\)] (b) -- [boson, rmomentum'=\(p_1\)] (f1),
			(b) -- [boson, rmomentum'=\(p_2\)] (c)		};
	\end{feynman}
\end{tikzpicture}
\begin{tikzpicture}
	\begin{feynman}
		\vertex (b);
		\vertex [above left=of b] (a1) {\(A_\alpha\)};
		\vertex [below left=of b] (a2) {\(A_\beta\)};
		\vertex [above right=of b] (b1) {\(W_\mu^-\)};
		\vertex [below right=of b] (b2) {\(W_\nu^+\)};
		\diagram* {
			(a1) -- [boson, momentum'=\(p_1\)] (b) -- [boson, rmomentum=\(p_2\)] (a2),
			(b1) -- [boson, momentum'=\(p_3\)] (b) -- [boson, rmomentum=\(p_4\)] (b2) };
	\end{feynman}
\end{tikzpicture}
%
%\noindent
%with the following vertex factors:
%
\begin{eqnarray}
 \mathcal{V}_{\mu\nu\lambda}^{(\gamma WW)} & = & i e \left[ \eta_{\mu \nu} (p_1 \! - \! p_2)_\lambda + \eta_{\nu \lambda} (p_2 \! - \! p_3)_\mu + \eta_{\lambda \mu} (p_3 \! - \! p_1)_\nu  \right]  \nonumber \\
& + & e \bigg[ \eta_{\mu \lambda} (p_2 \cdot p_3) \rho_\nu + p_{3\nu}  p_{1\lambda} \rho_{\mu}  \nonumber \\
& - &  p_{3\mu}  p_{2\lambda}  \rho_{\nu} - \eta_{\nu\lambda} (p_1 \cdot p_3) \rho_{\mu}  \nonumber \\
& - &  p_{1\nu}  p_{2\lambda} \rho_\mu + \eta_{\nu \lambda} (p_1 \cdot p_2) \rho_\mu  \nonumber \\
& + & p_{1\lambda} p_{2\mu} \rho_{\nu} - \eta_{\mu \lambda} (p_1 \cdot p_2) \rho_{\nu}  \nonumber \\
& + &  m_W^2 \left(\eta_{\nu \lambda} \rho_\mu - \eta_{\mu \lambda} \rho_\nu \right)  \bigg] \, , \label{vert_AWW} \\
\mathcal{V}_{\mu\nu\alpha\beta}^{(\gamma\gamma WW)} &=& i e^2 \left(\eta_{\mu \alpha} \eta_{\nu \beta} + \eta_{\mu \beta} \eta_{\nu \alpha} - 2 \, \eta_{\mu \nu}\eta_{\alpha \beta} \right) \nonumber \\
&+&  e^2 \left[ \left(\eta_{\mu \alpha} p_{4\beta} + \eta_{\mu \beta} p_{4\alpha} - 2 \eta_{\alpha \beta} p_{4\mu}  \right) \rho_{\nu}
 \right.
\nonumber \\
&+&\left. \left( \eta_{\nu \alpha} p_{3\beta} + \eta_{\nu \beta} p_{3\alpha} - 2 \, \eta_{\alpha \beta} p_{3\nu}  \right) \rho_{\mu} \right] \; . \label{vert_AAWW}
\end{eqnarray}

%%%%%%%%%%%%%%%%%%%%%%%%%
\section{Cross section for $e^+ \, e^- \rightarrow Z \, H$ with a generic LSV background} \label{app_B}
\indent

In the main text we have considered only the purely time-like case explicitly, as this simplifies the discussion leading to experimental bounds. Here we discuss how the results would change if we assume a general background containing both time- and space-like components.

As noted in sec.~\ref{sec_eeZH}, the LSV modification turns out to be a multiplicative factor, $\Pi(s, \rho)$. Assuming a generic 4-vector $\rho^\mu = (\rho_0, {\bm \rho})$, we have ($s = E_{\rm cm}^2$)
\begin{eqnarray}
\Pi(s, \rho) & = & 1 +  \left[\rho \cdot \left(p_1 + p_2\right) \right]^2 +  \left[\rho \cdot \left(p_1 +  p_2 - p_3\right)  \right]^2 \nonumber \\
& = & 1 + \rho_0^2 s + \left( \rho_0 E_4 -  {\bm \rho} \cdot {\bf p}_4 \right)^2 \, ,
\end{eqnarray}
but in the CM we have $E_4 = \sqrt{s} - E_3$ and ${\bf p}_4 = -{\bf q}$, with $E_3$ and $|{\bf q}|$ defined in eqs.~\eqref{eq_E3} and~\eqref{eq_modq}, respectively. With this, we have
\begin{equation}  \label{PI_spacelike}
\Pi(s, \rho) = 1 + \rho_0^2 s + \left[ \rho_0 \left( \sqrt{s} - E_3 \right) +  {\bm \rho} \cdot {\bf q} \right]^2 \, .
\end{equation}
Only the second line from eq.~\eqref{dcs_eeZH} carries angle-dependent factors, so we may concentrate only on this piece for the rest of this section. In fact, we may write
\begin{equation} \label{eq_int_pi}
\sigma_{eeZH}(\rho) \sim \int \left[ 1 + \frac{1}{m_Z^2} \left(E_3^2 - \vert {\bf q} \vert^2 \cos^2\theta\right)\right] \, \Pi\left( s, \rho  \right) d\Omega \, .
\end{equation}

\hfill \break
\hfill \break

This result can be applied to constrain the LSV 4-vector. Setting $\sigma_{eeZH}(\rho = 0) = \sigma_{eeZH}^{\rm SM}$, the sensitivities can be found by imposing (cf. eq.~\eqref{eq_lim_exp_eeZH})
\begin{equation} \label{cond_lim_eeZH}
\Bigg| \frac{\sigma_{eeZH} (\rho) - \sigma_{eeZH}(\rho = 0)}{\sigma_{eeZH}(\rho = 0)} \Bigg| \lesssim \delta\sigma^{\rm exp} \, .
\end{equation}

An important remark is in order. The calculations leading to the (differential) cross sections are performed in the CM frame, in which the $z$-axis is aligned with the initial beam direction --  the polar angle~$\theta$ is the deviation of the final particles from it. The 3-momentum ${\bf q}$ has the following components in the CM
\begin{equation}
{\bf q} =  |{\bf q}| \left( \sin\theta \cos\phi, \sin\theta \sin\phi, \cos\theta \right) \, ,
\end{equation}
but we must also determine the components of the LSV background in this frame. Note that we are dealing with three reference frames: the CM frame as defined above, the standard LAB frame ($z$-axis pointing to the local vertical) and the SCF ($z$-axis parallel to Earth's rotational axis). These are related via rotations and boosts, generally bringing time-dependent factors~\cite{ref_sun}.

As mentioned at the end of sec.~\ref{sec_intro}, the time component is approximately identical in the SCF and in the standard LAB frame. Given that the latter and the CM frame are related by a time-independent (spatial) rotation, the time component is the same in all three frames. The spatial components, on the other hand, are different in these frames, being related by rotation matrices: LAB$\leftrightarrow$SCF is time dependent~\cite{ref_sun}, whereas LAB$\leftrightarrow$CM is time independent. The LSV 4-vector is assumed to be approximately fixed in the SCF (which is inertial in all relevant time scales), but its components would seem to vary periodically in the other two frames.

For the sake of simplicity, however, we shall henceforth assume an ``instantaneous" LSV background in which the spatial components are time independent. This means that the scalar product in eq.~\eqref{PI_spacelike} becomes
\begin{equation}
{\bm \rho} \cdot {\bf q} = |{\bf q}| \left( \sin\theta \cos\phi\rho_x + \sin\theta \sin\phi\rho_y + \cos\theta\rho_z \right) 
\end{equation}
with $\{ \rho_x, \rho_y, \rho_z \}$ being fixed parameters in the CM. Plugging the equation above into eq.~\eqref{PI_spacelike} we are able to evaluate eq.~\eqref{eq_int_pi}. Inserting the result into eq.~\eqref{cond_lim_eeZH}, we find
\begin{equation} \label{F_G}
\big| F(s) \rho_0^2 + G(s, {\bm \rho}) \big| \lesssim \delta\sigma^{\rm exp} \, ,
\end{equation}
which relates the time and spatial components of the background to the experimental sensitivity. The functions $F(s)$ and $G(s, {\bm \rho})$ are given by
\begin{eqnarray}
F(s) & = & \frac{1}{20s} \frac{1}{[s - (m_{H}^2 - m_{Z}^2)]^2+8sm_{Z}^2} 
\nonumber \\
&&
\hspace{-1cm}
\times \, [ \, 25 s^4 - 40 s^3 m_{H}^2 + 10 s^2 m_{H}^4
\nonumber \\
&&
\hspace{-1cm}
+ 240 s^3 m_{Z}^2 
+ 60 s^2 m_{H}^2 m_{Z}^2 - 70 s^2 m_{Z}^4 
\nonumber \\
&&
\hspace{-1cm}
+ 40 s m_{Z}^2 (m_{H}^2 - m_{Z}^2)^2 + 5 (m_{H}^2 - m_{Z}^2)^4 \, ] \; ,
\end{eqnarray}

\begin{eqnarray}
G(s, {\bm \rho}) & = & \frac{1}{20s} \frac{1}{[s - (m_{H}^2 - m_{Z}^2)]^2+8sm_{Z}^2} 
\nonumber \\
&&
\hspace{-1cm}
\times \, [ \, 2 s^4 {\bm \rho}^2 - 8 s^3 m_{H}^2 {\bm \rho}^2 + 12 s^2 m_{H}^4 {\bm \rho}^2 
\nonumber \\
&&
\hspace{-1cm}
+12s^3 m_{Z}^2 {\bm \rho}^2 - 32 s^2 m_{H}^2 m_{Z}^2 {\bm \rho}^2 
\nonumber \\
&&
\hspace{-1cm}
-28 s^2 m_{Z}^4 {\bm \rho}^2 
+ 12 s m_{Z}^2 (m_{H}^2 - m_{Z}^2)^2 {\bm \rho}^2 
\nonumber \\
&&
\hspace{-1cm}
+2(m_{H}^2 - m_{Z}^2)^4 {\bm \rho}^2 
- s^4 \rho_{z}^2 
+ 4s^3 m_{H}^2 \rho_{z}^2 
\nonumber \\
&&
\hspace{-1cm}
-6s^2 m_{H}^4 \rho_{z}^2 + 4 s^3 m_{Z}^2 \rho_{z}^2 -
4 s^2 m_{H}^2 m_{Z}^2 \rho_{z}^2 
\nonumber \\
&&
\hspace{-1cm}
+4 s m_{Z}^2 (m_{H}^2 - m_{Z}^2)^2 \rho_{z}^2 
\nonumber \\
&&
\hspace{-1cm}
- 6 s^2 m_{Z}^4 \rho_{z}^2 
- (m_{H}^2- m_{Z}^2)^4 \rho_{z}^2 
\nonumber \\
&&
\hspace{-1cm}
-4 s m_{H}^2 (m_{H}^2 - m_{Z}^2)^2 (2 {\bm \rho}^2 - \rho_{z}^2) \, ] \; .
\end{eqnarray}

\begin{figure}[t!]
\begin{minipage}[b]{1.\linewidth}
\includegraphics[width=\textwidth]{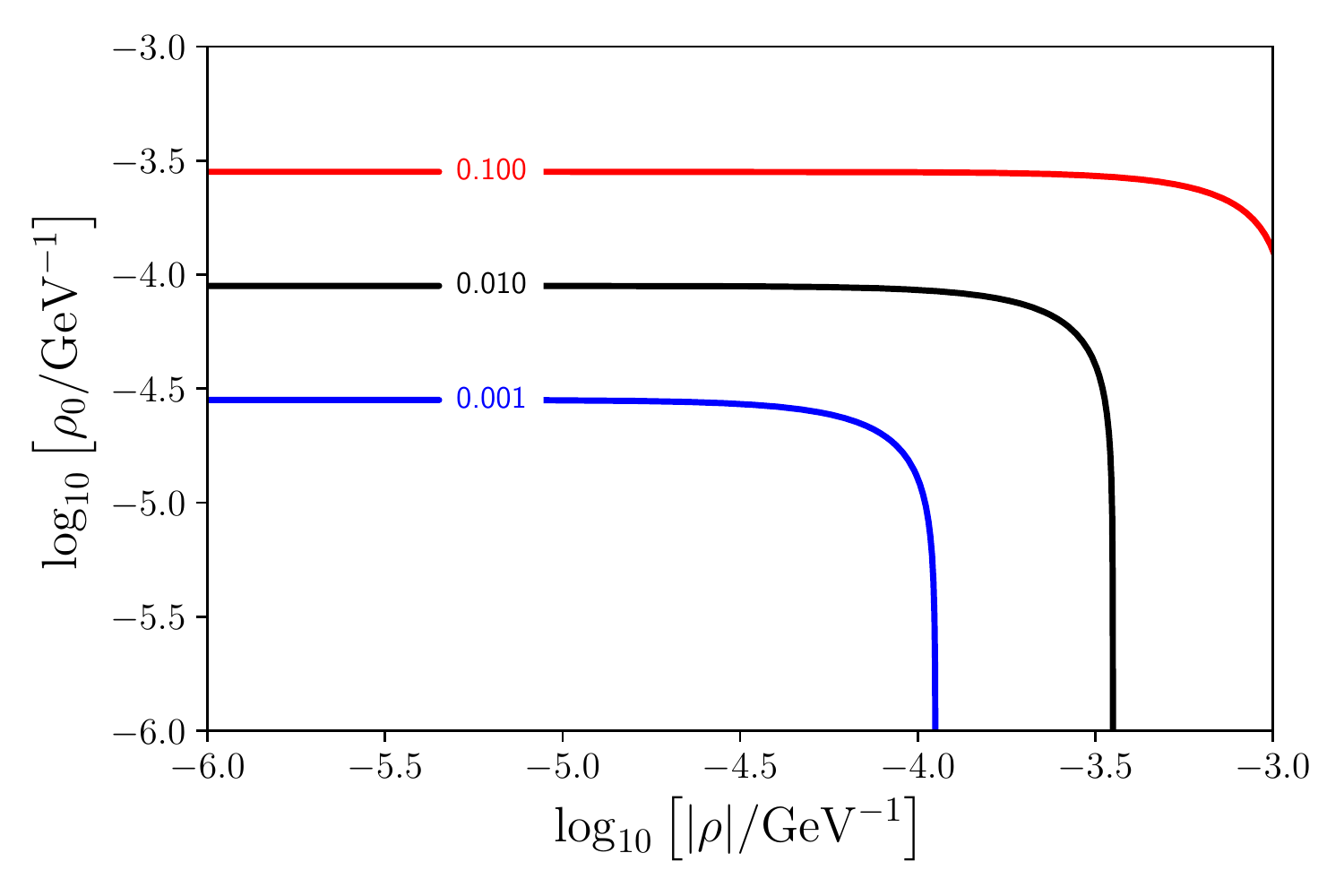}
\end{minipage} \hfill
\caption{Projected sensitivities for $\rho_0$ and $|{\bm \rho}|$ for the process $e^+ \, e^- \rightarrow Z \, H$. The curves are drawn for $\delta\sigma^{\rm exp} =$~0.1\%, 1\% and 10\%, in blue, black and red, respectively, cf. eq.~\eqref{eq_lim_exp_eeZH}. Here we have set $\sqrt{s} = 1$~TeV and assumed, for the sake of illustration, all spatial background components as equally large, {\it i.e.}, $|{\bm \rho}| = \sqrt{3} \rho_z$. Equation~\eqref{lim_eeZH} is dully recovered in the limit of vanishingly small $|{\bm \rho}|$ (see black curve).}
\label{fig_level_eeZH}
\end{figure}

It is worth noting that the components only appear quadratically and without mixing. This indicates that a non-trivial interplay can only be expected in regions where $|\rho_0|\approx |{\bm \rho}|$. This behavior is confirmed by the contours shown in fig.~\ref{fig_level_eeZH} in which different choices of the experimental sensitivity ($\delta\sigma^{\rm exp}$) are considered. As expected, the bounds for one component quickly become independent of the other when the relative magnitudes are slightly different. Moreover, for small $|{\bm \rho}|$, the black curve stabilizes at $\rho_0 \simeq 9 \times 10^{-5} \, {\rm GeV}^{-1}$, cf. eq.~\eqref{lim_eeZH}.

The comments made above justify our decision to focus on a purely time-like LSV 4-vector. Since we are primarily concerned with the order of magnitude of the sensitivities, this simpler case suffices and a more thorough discussion of a generic background does not bring essential information.

\end{document}